\documentclass[manuscript]{acmart} 
\AtBeginDocument{%
  }

\setcopyright{acmlicensed}
\copyrightyear{2026}
\acmYear{2026}
\acmDOI{XXXXXXX.XXXXXXX}
\acmConference[Conference acronym 'XX]{Make sure to enter the correct
  conference title from your rights confirmation email}{June 03--05,
  2018}{Woodstock, NY}
\acmISBN{978-1-4503-XXXX-X/2018/06}

\usepackage{xspace}
\newcommand{\sysname}{\textsc{PolyGandr}\xspace}

\usepackage{xcolor}
\usepackage{subcaption} 
\usepackage{pgffor} 

\usepackage{wrapfig} 
\usepackage{tcolorbox} 
\usepackage{hyperref}

\begin{document}


\title{\sysname: Steering Generative Models via Spatial Navigation through Latent Space} 

\author{Mingwei S.G. Li}
\correspondingauthor
\authornote{Both authors contributed equally to this research.}
\email{mingwei.li@tufts.edu}
\orcid{0000-0002-0457-8091}

\author{Susie S.Y. Li}
\authornotemark[1]
\email{suyang.li@tufts.edu}
\orcid{0009-0000-6375-182X}

\affiliation{%
  \institution{Tufts University}
  \city{Medford}
  \state{MA}
  \country{USA}
}

\author{Isabel Moreira de Oliveira}
\email{imdo@princeton.edu}
\orcid{0000-0003-0499-1617}

\author{Ryan P. Adams}
\email{rpa@princeton.edu}
\orcid{0000-0002-5704-6654}

\affiliation{%
  \institution{Princeton University}
  \city{Princeton}
  \state{NJ}
  \country{USA}
}




\author{Remco Chang}
\email{remco@cs.tufts.edu}
\orcid{0000-0000-0000-0000}

\affiliation{%
  \institution{Tufts University}
  \city{Medford}
  \state{MA}
  \country{USA}
}

\renewcommand{\shortauthors}{Li and Li et al.}

\begin{abstract}
Generative models support content creation across text, images, videos, and design disciplines, yet their interfaces rely primarily on natural language. While effective for specifying high-level intent, language can be inefficient and imprecise for fine-grained control, iterative refinement, and open-ended exploration. We present \sysname, a visual interface for exploring relationships between latent representations and generated outputs, complementing natural-language model steering. \sysname formulates generative control as high-dimensional navigation, enabling users to define starting points, steering directions, and step sizes through a unified map-and-mixing-palette interface. Combining global exploration with local refinement supports both broad design-space navigation and fine-grained output manipulation without repeatedly translating visual intent into prompts. We evaluate \sysname across three use cases: StyleGAN2 face generation, Stable Diffusion chair design, and rotational kirigami engineering. A formative study indicates that \sysname complements existing workflows by supporting real-time editing while simultaneously providing an overview and means to explore the broader design space.
\\

\end{abstract}


\begin{teaserfigure}
    \includegraphics[width=1\linewidth]{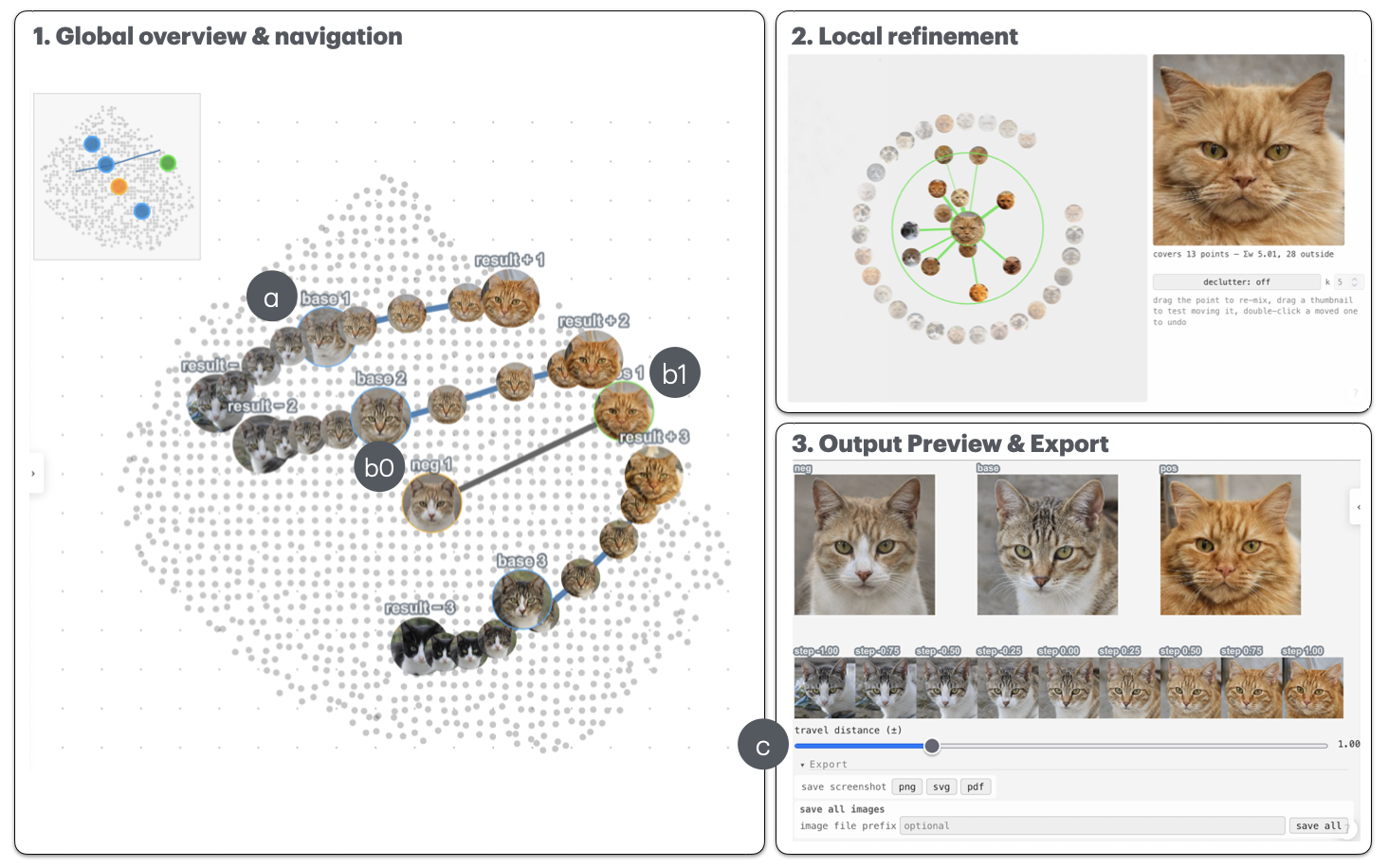}
    \caption{\sysname showcases how every cat has a path to becoming an orange cat \textasciicircum-\textasciicircum. By specifying (a) a starting point, (b0,b1) a steering direction from a contrastive pair, and (c) a step size, \sysname steers the model output in the global view (left, 1) from the base point toward the desired edit. The user can then locally refine the result (top right, 2) and export it in the desired format (bottom right, 3).}
  \Description{\sysname interface showing overview map-like explorer on the left and the output of a generative model in this case image of a cat on the right}
  \label{fig:teaser}
\end{teaserfigure}


\maketitle

\section{Introduction}

Generative models support content creation across text, images, videos, and design disciplines, yet interfaces for steering or editing their outputs rely primarily on natural language~\cite{podell2023sdxlstablediffusion, rombach2022highresolutionimagesynthesislatent}. 
Exploring these outputs requires users to discover possibilities and refine promising results. A user may recognize a desirable variation in an example without knowing how to express it as a prompt or parameter~\cite{bo2026steerablechatbotsexploringpersonalization}.


Prior interfaces support exploring generated examples, discovering and refining semnatic directions, and editing individual outputs. Building on these capabilities, we investigate how an interface can connect the discovery of a useful contrast with inspection and revision of the resulting transformation. 

We present \sysname, a visual interface that connects example-based steering with \textbf{global exploration} and \textbf{local refinement}. Users specify a starting point, define a direction through contrasting examples, and adjust a step size to control the strength of the transformation. A global map displays sampled outputs and projects steering trajectories, and a linked local mixing view supports users to adjust the examples contributing to the starting point or either endpoint of the direction. Together, these views support an iterative workflow of constructing, inspecting, and revising steering operations. 

We demonstrate \sysname with StyleGAN2 face and cat generation, and Stable Diffusion XL chair generation. We also explore its extension to a parametric engineering design space through an exploratory user study with a domain expert. The demonstrations illustrate the steering workflow, while the user study identifies opportunities for combining broad exploration with domain-specific editing and highlights the need for guidance when designs become infeasible. 

Specifically, we contribute the following: 
\begin{itemize}
    \item An interaction framework for steering generative models by connecting global exploration with local refinement. Users construct a steering  operation from a base point, contrasting examples, and a step size, then inspect and revise the examples defining that operation. 
    \item \sysname, an interactive system that instantiates this framework through a linked global overview and local mixing interface. The system visualizes steering trajectories in the overview and allows users to reposition neighboring examples to adjust their influence on the base point or either endpoint of a steering direction.\footnote{The code will be made publicly available upon acceptance of the paper. \url{https://placeholder.github.url}}
    \item Demonstrations with StyleGAN2 and Stable Diffusion XL, and a formative engineering case study involving a rotational-kirigami expert. These applications illustrate the approach across generative and parametric design spaces and identify opportunities for domain-specific refinement and feasibility guidance.
\end{itemize}

\section{Related Work}\label{sect:related-work}

We review two areas of related work: model steering and exploration. The first covers steering mechanisms, computational techniques, and their theoretical foundations. The second focuses on interfaces that draw on these methods to support human-in-the-loop interactive steering and exploration.


\subsection{Model Steering and Conditioning Methods}

Generative-model steering can operate on latent representations, intermediate features, model parameters, conditioning signals, or the generated output itself. Early feature or latent space methods manipulate model's intermediate representations directly. 
Deep Feature Interpolation~\cite{upchurch2017deepfeatureinterpolationimage} derives attribute directions from differences between groups of examples, while GANSpace~\cite{harkonen2020ganspace}, InterFaceGAN~\cite{InterFaceGAN}, SeFa~\cite{sefashen2021}, and StyleSpace~\cite{wu2020stylespace} identify interpretable directions in GAN representations. 
LatentCLR~\cite{yüksel2021latentclr} identifies interpretable directions in GAN via contrastive learning.
StyleCLIP~\cite{patashnik2021styleclip} further connects such manipulation to natural-language specifications. Other methods provide more spatially explicit control: EditGAN~\cite{ling2021editgan} supports semantic image editing through segmentation-based manipulation, while DragGAN~\cite{Pan2023draggan} enables direct point-based manipulation on the generative image manifold.
These results motivate our use of simple vector arithmetic while also highlighting that linear steering should be understood as an approximation whose validity may vary across the latent space and model architecture.

Diffusion models support an even broader range of control mechanisms. Prompt-to-Prompt~\cite{hertz2022prompttopromptimageeditingcross} manipulates cross-attention to preserve image structure while modifying textual concepts, while SDEdit~\cite{meng2022sdedit} uses an input image as guidance by partially noising and denoising it under a new condition. InstructPix2Pix~\cite{brooks2023instructpix2pix} instead learns image editing directly from natural-language instructions. ControlNet~\cite{zhang2023controlnet} and T2I-Adapter~\cite{mou2023t2iadapter} introduce explicit spatial conditions such as pose, depth, edges, and segmentation. Other methods provide semantic guidance through attention or diffusion trajectories~\cite{chefer2023attendandexcite,brack2023segasemanticguidance,bartal2023multidiffusion}.
Continuous semantic controls can also be constructed by adapting model parameters. Concept Sliders~\cite{gandikota2023conceptsliders} learns low-rank parameter directions using LoRA~\cite{hu2021lora}, while SliderSpace~\cite{gandikota2025sliderspace} discovers multiple interpretable slider directions from a single prompt. AdaptiveSliders~\cite{gupta2025adaptivesliders} further considers how such controls can be aligned with user intent. These methods provide increasingly powerful mechanisms for defining \emph{what} can be controlled. \sysname is complementary: rather than introducing another mechanism for learning a semantic direction, we focus on how users spatially construct, inspect, combine, and refine steering operations through examples.

\subsubsection{Geometry of Generative Latent Spaces}

Latent space steering often assumes that meaningful changes can be represented by linear directions in a locally Euclidean space. However, generative models generally can induce non-uniform and curved geometries. Earlier work formalized this through a pullback of Riemannian metrics induced by generative mappings, enabling geodesic interpolation and analysis of local distortion~\cite{arvanitidis2018latentoddity, shao2018riemannian}.
More recent work extends this geometric view to diffusion and foundation models, showing that latent geometry can vary across timesteps and conditions, exhibit non-uniform local scaling and rank, and depart from linearity across different regions of the space~\cite{park2023diffusiongeometry,humayun2025manifolds,pmlr-v235-hahm24a}. Other studies further relate generative behavior to manifold curvature, tubular neighborhoods, score-Jacobian spectra, and locally linear geodesic structure that can break down across phase boundaries~\cite{pmlr-v251-sakamoto24a,ventura2025geometric,pmlr-v267-lobashev25a}.

Together, these results suggest that the validity of linear latent space steering is inherently local and model dependent. A direction that is meaningful around one point, timestep, or conditioning context may not remain geometrically or semantically consistent elsewhere. \sysname therefore treats linear vector arithmetic as a simple and fast approximation. With this simplification we encourage users to interactively explore and refine, to expose where this approximation succeeds or breaks down.

\subsection{Interactive Exploration and Steering Interfaces}

A parallel line of work studies how users interact with generative spaces once controllable representations are available. GANSlider~\cite{GANSlider} exposes multiple semantic GAN directions (via PCA) through sliders and studies the effects of feedforward previews in film-strip like interface. GANzilla~\cite{Evirgen_2022_GANzilla} allows users to discover directions iteratively from selected examples, while GANravel~\cite{EvirgenGANravel_2023} supports refining and disentangling those directions. SwipeGANSpace~\cite{Nakashima_2024_swipeganspace} similarly supports efficient latent space exploration through pairwise comparison interactions. These systems show the value of allowing users to discover and refine controls rather than requiring all semantics to be specified in advance.

Other systems emphasize broader exploration and creative navigation. SpaceSheets~\cite{Loh2018spacesheet} organizes latent space exploration through a spreadsheet-like interface, while Browsing the Latent Space~\cite{browsinglatentspace} treats latent representations as a space for interactive design exploration. GANCollage~\cite{wan_gancollage_2023} uses generative models as an exploratory mood-board medium for ideation. LatentZoom~\cite{Li2026LatentZoom} supports multiscale visual exploration of local behavior in generative latent spaces. Latent Space Map~\cite{zhang2025latentspacemap} similarly projects generative representations into a 2D map to support visual exploration and help users locate regions containing desirable generated outputs. While this work focuses primarily on locating and selecting promising samples, \sysname uses the global map as an interaction surface for constructing, visualizing, and refining steering operations, generating unseen data with example-based local manipulation.
Together, these systems motivate treating a generative representation not only as a parameter space to optimize, but also as a landscape that can be browsed and inspected.

For diffusion-based generation, PromptPaint~\cite{Chung_2023PromptPaint} is particularly relevant to our interaction framing: it uses a paint-mixing metaphor to combine text prompts and supports continuous exploration between semantic specifications. PromptCharm~\cite{Wang_2024_promptcharm} similarly combines multiple modalities for iterative prompt construction and refinement. These interfaces move beyond a single text box toward richer spatial and compositional interaction with generative models.

\sysname builds on these directions but focuses on a different unit of interaction: \emph{navigation through example-defined vector arithmetic}. Users select or interpolate a base point, construct a steering direction from contrasting examples, vary the step size, and inspect the resulting trajectory in a global latent space overview. Local mixing further allows each component of this algebra to be refined through neighboring examples. Rather than assuming a predefined editing objective, \sysname emphasizes free exploration: users can browse latent space structure, discover contrasts, construct directions, and observe how those directions behave across different regions. This complements sliders and prompt-based controls, which are particularly effective when the desired semantic manipulation is already known and explicitly parameterized.

\section{Interface Design Goals and Solution Guidelines}

This section presents three design goals (DG) that structure \sysname around how users reason about and navigate generative spaces. We first support understanding the model's global range of outputs, then enable precise refinement within a selected region, and finally maintain a consistent interaction metaphor across these two scales. For each goal, we describe the underlying user need and the corresponding design solution guideline (SG) implemented in \sysname.


\subsection{DG1: Support Orientation within the Sampled Design Space}
Users need to understand what a generative model can produce under a given set of conditions before they can effectively decide where to explore, what to refine, and how to divide work between themselves and the model. This requires reasoning across many possible outputs rather than inspecting isolated generations one at a time. For example, when generating images of a particular subject or style, users may want to understand the overall range of forms, materials, compositions, or other variations that remain available under that conditioning. Such an understanding helps users identify both the model's capabilities and its limitations, and informs which aspects are better specified directly by the user versus left open for model-driven exploration.

\paragraph{SG1:} In response, \sysname tightly couples this need for understanding output coverage with a global overview of the generative landscape. The interface presents sampled model outputs in a map-like view, allowing users to recognize broad patterns, clusters, transitions, and gaps across the space. Rather than limiting exploration to the initially sampled points, users can place and move lenses over regions of interest, inspect neighborhoods at different spatial scales, and generate additional samples on demand around a selected region. This supports a progressive workflow in which users first form a broad mental model of the generative space, then zoom into regions that appear relevant or promising for more detailed exploration.

\subsection{DG2: Support Precise Refinement of Examples and Steering Directions}

Once users identify a promising region or output in the global landscape, they need to refine it according to preferences, constraints, or goals that may not have been fully specified in advance. At this stage, the challenge shifts from broad exploration to precise local control: users need to understand what variations are available around the current output and make incremental changes without losing the characteristics that motivated their selection. Because the relevant directions of variation can depend on the current location in the generative space, these controls should adapt to the local context rather than rely only on globally predefined editing axes.

\paragraph{SG2:} In response, \sysname provides a local mixer centered on the current point. The system samples neighboring latent representations and exposes them through a color-mixing-palette-like interface, where users can continuously interpolate among nearby examples using weighted combinations. This allows users to inspect locally available variations, fine-tune the current output, and construct intermediate representations that better match their evolving intent while remaining anchored to the selected neighborhood.

Local refinement can take different forms across domains. It may use semantic directions in StyleGAN, LoRA-based diffusion controls, or natural-language and image-space conditioning. In addition to the color-mixing-palette-like interface, our framework also leaves room for generic local controls grounded in the data and domain context.
In domain-specific settings, refinement can be defined directly by meaningful parameters and constraints; for rotational kirigami design (\autoref{sect:eval}), for example, we use a polygonal hub editor together with controls for target height and twist.

\subsection{DG3: Promote Consistent Reasoning across Global and Local Navigation}
Users need to move fluidly between broad exploration and precise refinement without having to learn fundamentally different interaction models at each scale. 
In the global view, they may only know approximately where they want to go, using visible examples in the overview as landmarks rather than specifying an explicit direction local in a latent space. 
In the local view, they need finer control over nearby variations while preserving the same spatial intuition. Maintaining a consistent interaction metaphor across these scales aims to reduce the cognitive shift between exploration and refinement and helps users reason about both as navigation through the same generative space.

\paragraph{SG3:} In response, \sysname uses the same ``I want to go there'' interaction paradigm in both views. In the global view, the layout is fixed and users navigate toward existing examples or regions in the UMAP projection. In the local view, the neighboring examples remain spatial targets, but their layout can be directly \textit{rearranged} to reshape the interpolation surface. By repositioning these neighbors, users gain finer control over the relative influence of nearby examples and can, in principle, access any convex combination within the simplex spanned by the selected neighbors, given sufficient adjustment of the layout.

\section{Method} \label{sect:method}
We describe the core techniques underlying model output steering in \sysname.

\subsection{Three Components of Model Output Steering}
Many model-steering techniques rely on simple vector arithmetic: starting from one latent vector, we add or subtract a vector representing a semantic direction. Later in \autoref{sect:use-cases}, we examine how these operations enable meaningful transformations in real-world datasets and generative models.

\subsubsection{The Base $z_0$: Where Do We Start?}

As with navigating a map, latent space exploration begins from familiar landmarks rather than raw coordinates. Each sample $z_i=(z_{i1},\ldots,z_{iM})$ serves as such a landmark in the high-dimensional space.
A new point can then be defined as a weighted combination of nearby samples. Varying the weights continuously allows continuous navigation between known examples; smoothness and differentiability depend on the chosen weighting function.

\subsubsection{The Direction $\vec{d}$: Which Way Do We Go Next?
}
The direction specifies how a base point moves through latent space. More generally, it can be defined by any vector field, for example from local model geometry via a Jacobian basis. A simpler and cheaper alternative is to derive it from contrastive pairs.

As a simple example, a direction can be defined by subtracting two endpoints: an example representing the desired destination minus an example representing the opposite concept. In this case, the direction is specified using only two examples:
$$
\vec{d} = z_{\mathrm{pos}} - z_{\mathrm{neg}}
$$
A direction can also be defined as the average of the differences between matched positive-negative pairs:
$$\vec{d} = \operatorname{set\_difference} \left( \{z_{\mathrm{pos}1}, \ldots, z_{\mathrm{pos}K_1}\}, \{z_{\mathrm{neg}1}, \ldots, z_{\mathrm{neg}K_2}\} \right)$$
or as the difference between the averages of two groups:
\begin{align}
\vec{d} = \operatorname{average} \left( z_{\mathrm{pos}1}, \ldots, z_{\mathrm{pos}K_1} \right) - \operatorname{average} \left( z_{\mathrm{neg}1}, \ldots, z_{\mathrm{neg}K_2} \right) \label{eq:direction}
\end{align}

In general, steering directions can be viewed as a vector field defined over the latent space. 
In \sysname, we simplify this formulation by defining a direction through weighted averages over neighborhoods of examples, allowing finer-grained and potentially more disentangled steering via \autoref{eq:direction}. 
That is, given two groups representing opposite concepts (e.g., short versus long hair), we define a global steering direction as the element-wise vector difference between their mean latent representations. This contrast captures the variation separating the groups and can be applied to a new base point to steer it toward either concept.

Under this simplifying assumption, the resulting steering vector is applied globally to any base point, effectively treating the latent space as Euclidean. This simple formulation enables real-time interaction in \sysname that allows users to specify the base point, direction, and step size interactively.

\subsubsection{The Step Size $\alpha$: How Far Should We Go?}

Once we know which direction to move, the next question is how far to go. The step size controls the magnitude, or strength, of the change applied to the base point. A small step produces a subtle change, while a larger step pushes the result farther along the chosen direction. Let $\alpha$ denote the step size relative to the distance between the positive and negative examples.

\begin{align}
z_{\mathrm{new}} &= z_0 + \alpha \left( z_{\mathrm{pos}} - z_{\mathrm{neg}} \right)
\label{eq:algebra}
\end{align}

In \sysname, the step size $\alpha$ is controlled directly through a slider, determining how far the base point moves along the selected steering direction.

\begin{wrapfigure}{r}{0.4\textwidth}
  \centering
    \includegraphics[width=1\linewidth]{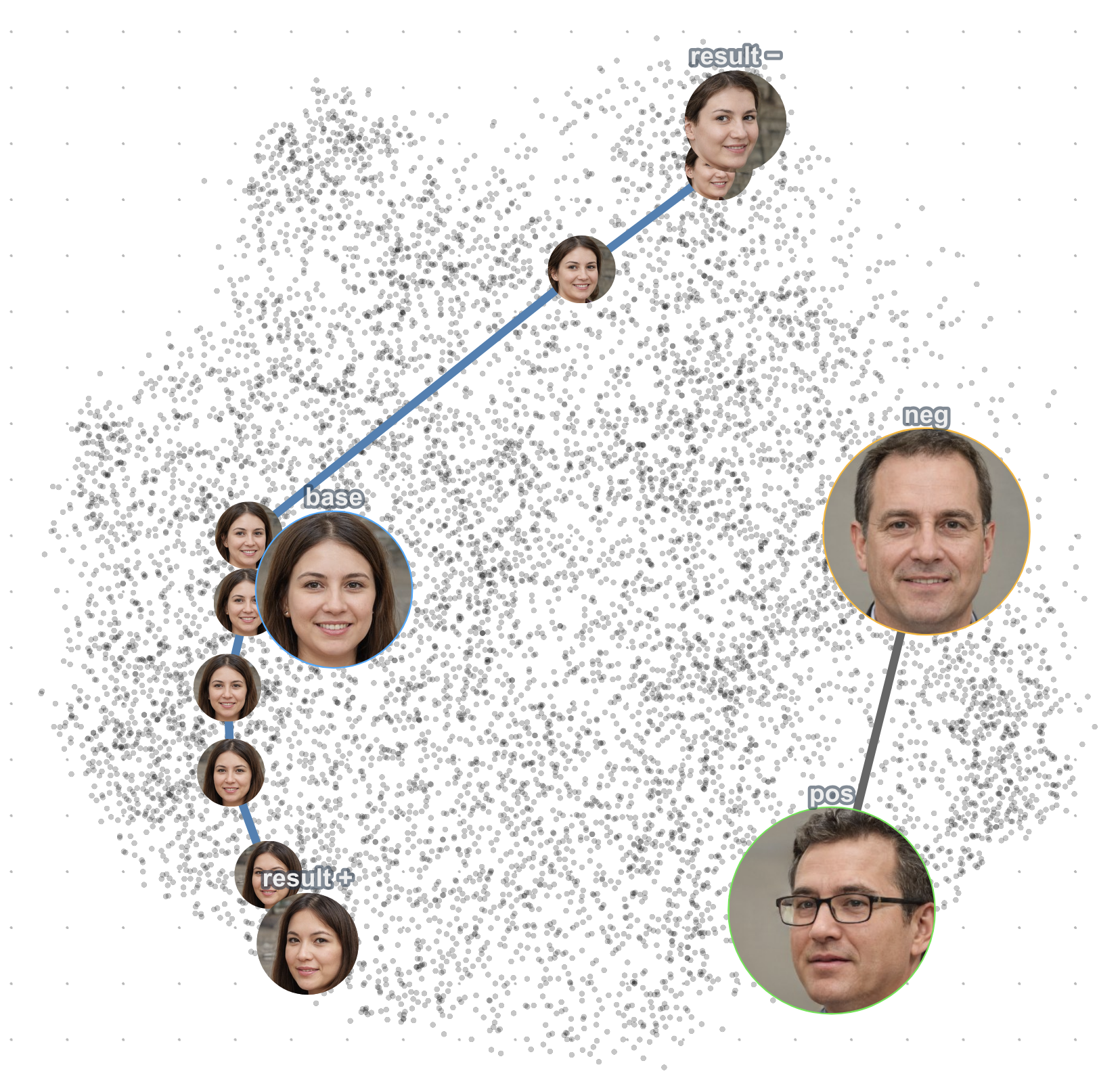}
    \caption{The global view project samples of generated examples with a parametric UMAP.}
    \Description{A global map of generated samples positioned by a two-dimensional parametric UMAP projection.}
        \label{fig:global-1}
\end{wrapfigure}

\subsection{Global Overview and Exploration via Zoom-in Lenses}

The global overview aims to expose the degrees of freedom and major modes of variation available in the generative process via a map of generated samples. 

Not all generative parameters need to remain globally explorable. Some can be fixed by user intent or domain constraints: for example, a prompt may constrain subject and style, while physical or structural constraints can filter out infeasible regions. Strongly semantic dimensions, such as age, pose, or expression, may instead be exposed through explicit controls such as sliders or menus. The global overview can therefore focus on residual, correlated, or previously unknown modes of variation that are difficult to parameterize a priori.

Once known variations are fixed, filtered, or binned, we project the remaining high-dimensional representations into 2D or 3D using dimensionality reduction such as UMAP. For example, for an unconditional generative model, such as a StyleGAN trained on human faces~\cite{karras2020analyzing, stylegan, ffhq}, we:

\begin{enumerate}
    \item Sample globally from the model's prior distribution.
    \item Choose an intermediate layer activation as the latent representation of the sample. Project the latent representation data into 2D or 3D using a parametric dimensionality reduction technique, such as parametric UMAP, so that it is capable of embeding new unseen samples.
    \item Use the interpolation scheme to-be-described to support a ``latent lens'' interface, for exploring, zooming into, and navigating regions in the global map view through distance-weighted combinations of nearby samples.
\end{enumerate}



\paragraph{The interpolation scheme}
Consider $K$ neighboring anchor points with latent codes
$\{z_1,\ldots,z_K\}$ and corresponding 2D positions
$\{p_1,\ldots,p_K\}$, where $p_i=(x_i,y_i)$. 
For any control-point
location $p=(x,y)$, we combine the latent codes of the reference samples,
$z_1,\ldots,z_K$, using a weighted linear combination:
\begin{align}
z(p) = \sum_{i=1}^{K} w_i(p) z_i,
\qquad \text{where we require}
\sum_{i=1}^{K} w_i(p) =1 \label{eq:weighted-sum}
\end{align}
where each weight $w_i$ is determined by the 2D distance
$d_i=\lVert p-p_i\rVert_2$ between reference point $p_i$ and the
control point $p$.
 The choice of distance-to-weight function is flexible. 
Consider three examples: Inverse-distance weighting $w^{inv}$, smooth step (down) $w^{smoothstep}$ and a shifted negative log $w^{neg\log}$:

\begin{align}
w_i^{\mathrm{inv}} &= \frac{1/d_i}{\sum_{j=1}^{K}1/d_j},
\label{eq:weight-function-1}
\\
w_i^{\mathrm{smoothstep}} &= \frac{\phi(d_i/R)} {\sum_{j=1}^{K}\phi(d_j/R)},
\qquad
\phi(t)=\begin{cases}
1-3t^2+2t^3, & t<1,\\
0, & t\geq1,
\end{cases}
\label{eq:weight-function-2}
\\
w_i^{neg\log} &= \frac{\psi(d_i/R)}{\sum_{j=1}^{K}\psi(d_j/R)},
\qquad
\psi(t)= \begin{cases}
-\log t, & t<1,\\
0, & t\geq1.
\end{cases}
\label{eq:weight-function-3}
\end{align}

Inverse-distance weighting has global support, so every reference
sample contributes at every location and the mixture changes gradually.
The smoothstep and logarithmic functions have compact support controlled
by the cutoff radius $R$, producing a sparser and more local mixture.
Smoothstep treats nearby points relatively uniformly and fades smoothly
to zero near the boundary, whereas logarithmic weighting maintains a
stronger preference for the nearest sample throughout its support.
Thus, the weighting function controls both the locality of the mixture
and how strongly the interface discriminates among nearby examples.

Through this interpolation, a simple 2D movement simultaneously changes
the contributions of multiple high-dimensional latent vectors. Unlike
a slider that follows a single predefined direction, the interface
allows users to continuously vary the relative influence of several
examples---analogous to mixing different proportions of paint.

These choices therefore encode different properties of the navigation surface. The weighting function determines how influence varies with distance; the cutoff radius determines the locality and sparsity of that influence; and normalization determines whether interpolation is restricted to convex combinations or may extrapolate beyond them. Rather than treating these choices as implementation details, they can be viewed as design parameters that determine the geometry of the 2D navigation surface induced in latent space.

\begin{figure}[t]
    \centering

    \begin{subfigure}[t]{0.32\textwidth}
        \centering
        \includegraphics[width=\linewidth]
        {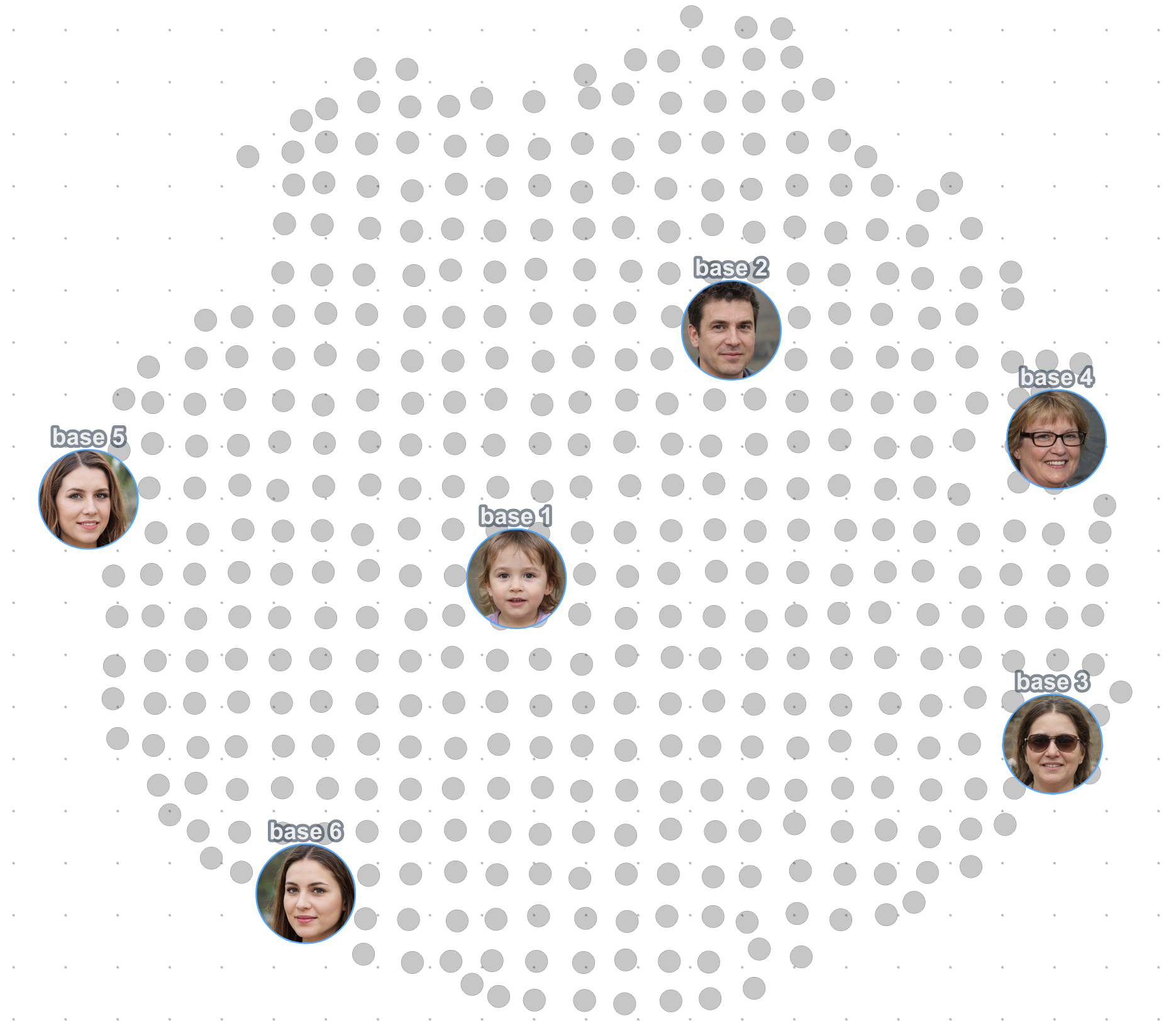}
        \caption{\textbf{Mode 1: Base Mode.}
        In the global view's base-only mode, users can add multiple
        base lenses to examine the overall latent space landscape,
        shown here with sample-point thumbnails.}
        \Description{A global latent space map with multiple base lenses and sample image thumbnails.}
        \label{fig:mode-1}
    \end{subfigure}
    \hfill
    \begin{subfigure}[t]{0.32\textwidth}
        \centering

        \includegraphics[width=\linewidth]
        {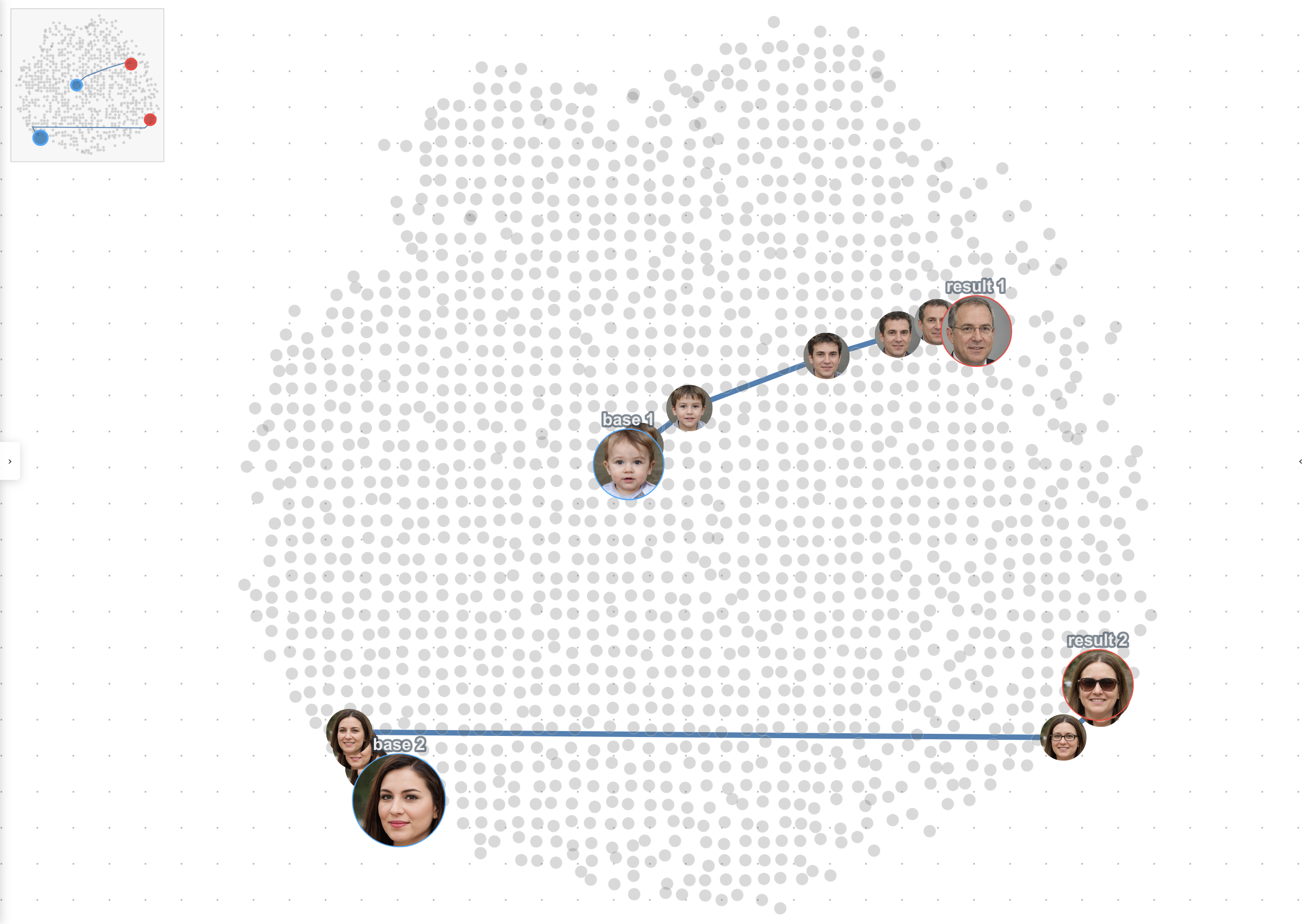}

        \vspace{2pt}

        %
            \includegraphics[width=0.137\linewidth]
            {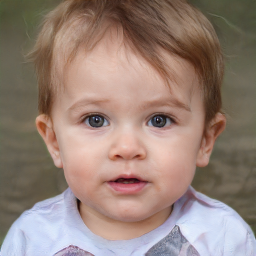}%
            \hfill
            \includegraphics[width=0.137\linewidth]
            {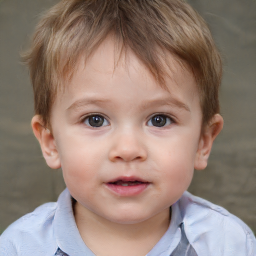}%
            \hfill
            \includegraphics[width=0.137\linewidth]
            {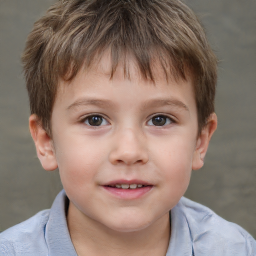}%
            \hfill
            \includegraphics[width=0.137\linewidth]
            {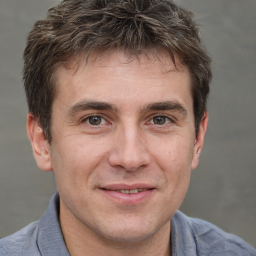}%
            \hfill
            \includegraphics[width=0.137\linewidth]
            {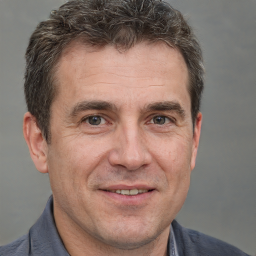}%
            \hfill
            \includegraphics[width=0.137\linewidth]
            {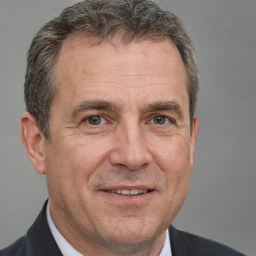}%
            \hfill
            \includegraphics[width=0.137\linewidth]
            {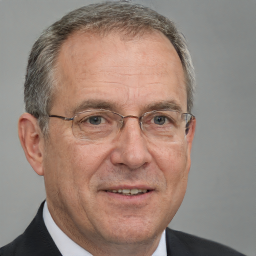}%
            \hfill

        \vspace{2pt}

        %
            \includegraphics[width=0.137\linewidth]
            {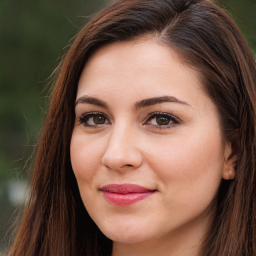}%
            \hfill
            \includegraphics[width=0.137\linewidth]
            {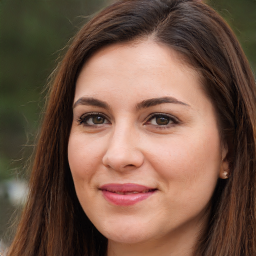}%
            \hfill
            \includegraphics[width=0.137\linewidth]
            {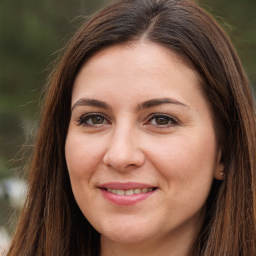}%
            \hfill
            \includegraphics[width=0.137\linewidth]
            {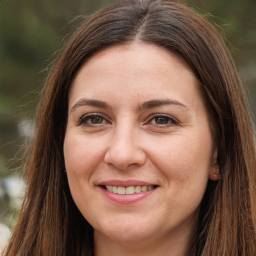}%
            \hfill
            \includegraphics[width=0.137\linewidth]
            {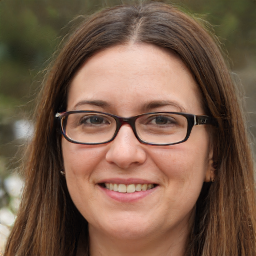}%
            \hfill
            \includegraphics[width=0.137\linewidth]
            {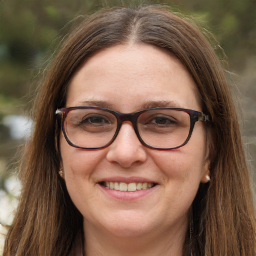}%
            \hfill
            \includegraphics[width=0.137\linewidth]
            {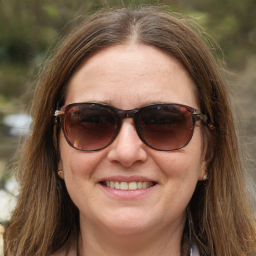}%
            \hfill

        \caption{\textbf{Mode 2: Pair Mode.}
        Pair mode can display multiple independent interpolation
        trajectories in the global view. Here, one trajectory connects
        a child to a man with eyeglasses, while the other connects a
        woman without sunglasses to a woman with sunglasses.}
        \Description{A global map above two rows of seven face images: a child transitions to a man with eyeglasses, and a woman without sunglasses transitions to a woman with sunglasses.}
        \label{fig:mode-2}
    \end{subfigure}
    \hfill
    \begin{subfigure}[t]{0.32\textwidth}
        \centering
        \includegraphics[width=\linewidth]
        {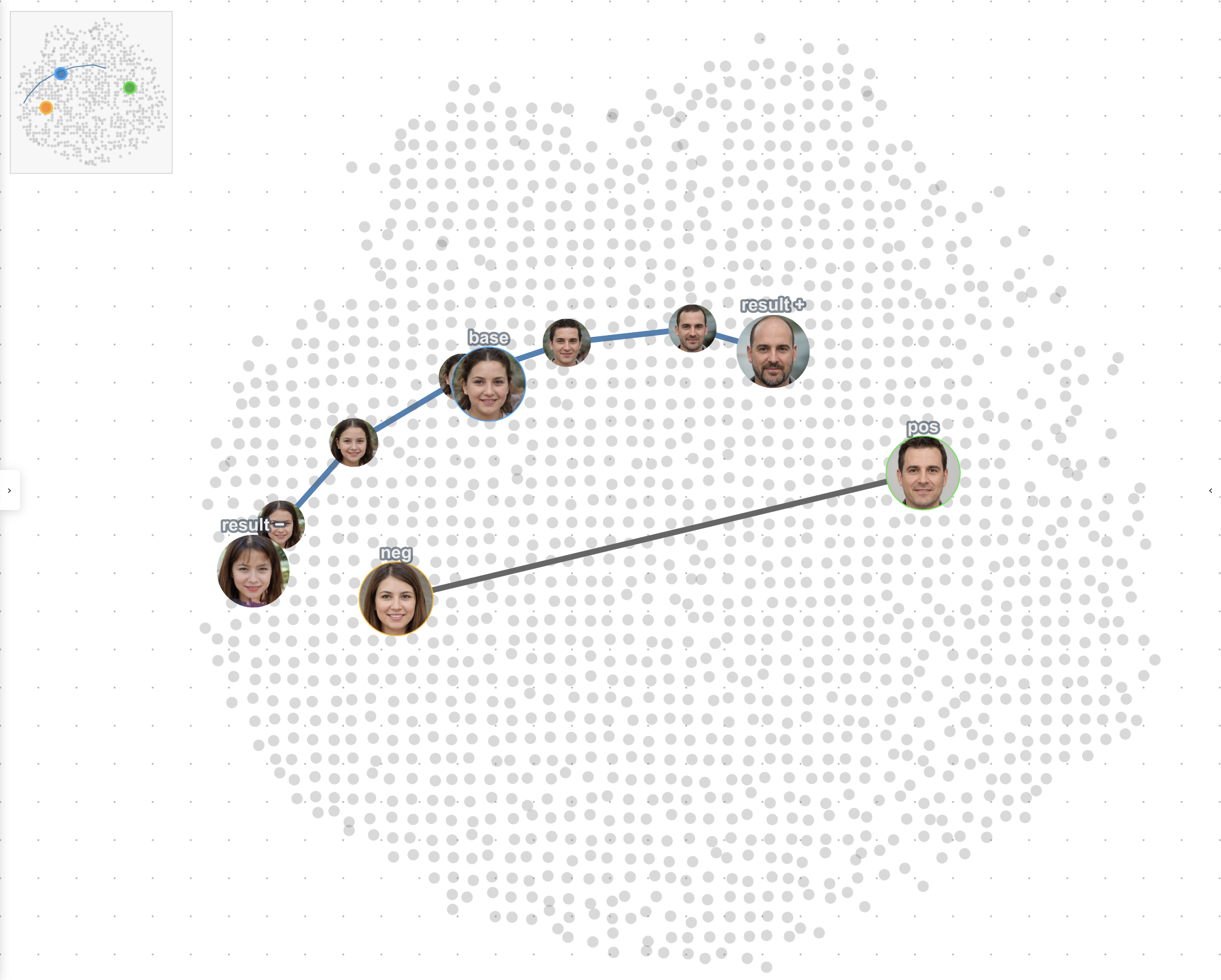}
        \caption{\textbf{Mode 3: Steer Mode.}
        \sysname supports interactive vector-arithmetic steering by
        letting users specify a base point, a contrastive direction
        (\textit{negative} and \textit{positive}), and a step size.
        Sweeping the step size reveals the full steering trajectory,
        which is mapped back into the global overview through parametric
        UMAP together with the final output.}
        \Description{A global overview displaying a steering trajectory and generated output, with controls for the base point, negative and positive examples, and step size.}
        \label{fig:mode-3}
    \end{subfigure}

    \caption{Global navigation modes in \sysname.
    Left: base-mode exploration with multiple lenses and optional
    image thumbnails. Middle: pair-mode interpolation with two
    independently specified trajectories and their generated outputs.
    Right: steer mode for vector-arithmetic manipulation using a base
    point, contrastive direction, and step size.}
    \Description{Three side-by-side global navigation modes in
    \sysname: base exploration, pair interpolation, and
    vector-arithmetic steering.}
    \label{fig:global-modes}
\end{figure}

\subsubsection{Three modes of exploration}

\paragraph{Mode 1: Base-only Mode}
In the simplest exploration mode, a base lens is a single draggable control that allows users to move across the DR plot and inspect local regions through the weighted interpolation described above. Users can add multiple base lenses to examine the overall latent space landscape, as shown in \autoref{fig:mode-1}.

\paragraph{Mode 2: Pair Mode}
To test an interpolation scheme, users can select pairs of lenses and uniformly interpolate between their endpoints over a specified number of steps. \autoref{fig:mode-2} shows two such paths defined by two pairs of endpoints.


\paragraph{Mode 3: Steer Mode. Interactive steering through vector arithmetic.}
Building on this global navigation space, \sysname uses the standard vector-arithmetic (\autoref{eq:algebra})
and provides interactive controls for specifying each of its three components: the base point $z_{\mathrm{base}}$, the steering direction $z_{\mathrm{pos}}-z_{\mathrm{neg}}$ defined by a contrastive positive-negative pair, and the step size $\alpha$ (\autoref{fig:mode-3}). Rather than showing only the final steered result, the interface also sweeps $\alpha$ over a range of values to visualize the full steering trajectory in latent space. Both the final output and the intermediate points along this trajectory are then mapped back into the global overview using the learned parametric UMAP, allowing users to see where the steering operation moves relative to the surrounding latent space structure.

\begin{figure}[h]
    \centering
    \includegraphics[width=0.4\linewidth]{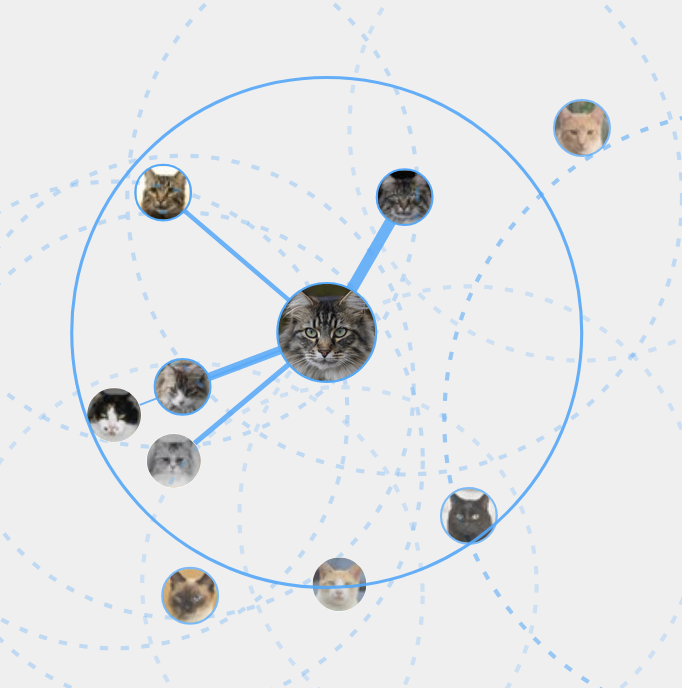}
    \caption{Local interpolation among cat latent codes. The range circle indicates the cutoff radius, while each neighboring sample within range contributes to the central control point according to its distance and the selected weighting function (\autoref{eq:weight-function-1}--\autoref{eq:weight-function-3}).}
    \Description{Cat examples surround a central interpolation control point, with a circle marking the radius within which neighboring examples contribute to the mixture.}
    \label{fig:local-mix-cat}
\end{figure}

\subsection{Local Refinement via Free-Form Sample Interpolation}
Local refinement uses the same interpolation paradigm as the global view (\autoref{eq:weighted-sum}), except that the neighboring nodes can now be freely repositioned (\autoref{fig:local-mix-cat}). This gives the user direct control over how the examples are mixed: based on their distances to the central control point, the interpolation can realize different convex combinations of the surrounding examples.

Fixing the anchors yields only a 2D parameterization and cannot generally cover the full latent space. With freely repositionable anchors and normalized nonnegative weights, interpolation is bounded by the anchors' convex hull, a $(K-1)$-simplex when they are affinely independent. For any fixed layout, however, the 2D interface only slices a 2D subset of this high dimensional space. We therefore should carefully consider the coverage of this surface, relative to the anchor-induced hull, the local data manifold, and the semantic variation of the generated outputs within the neighborhood. (See. \autoref{sect:discussion}).

\begin{wrapfigure}{r}{0.4\textwidth}
  \centering
    \includegraphics[width=1\linewidth]{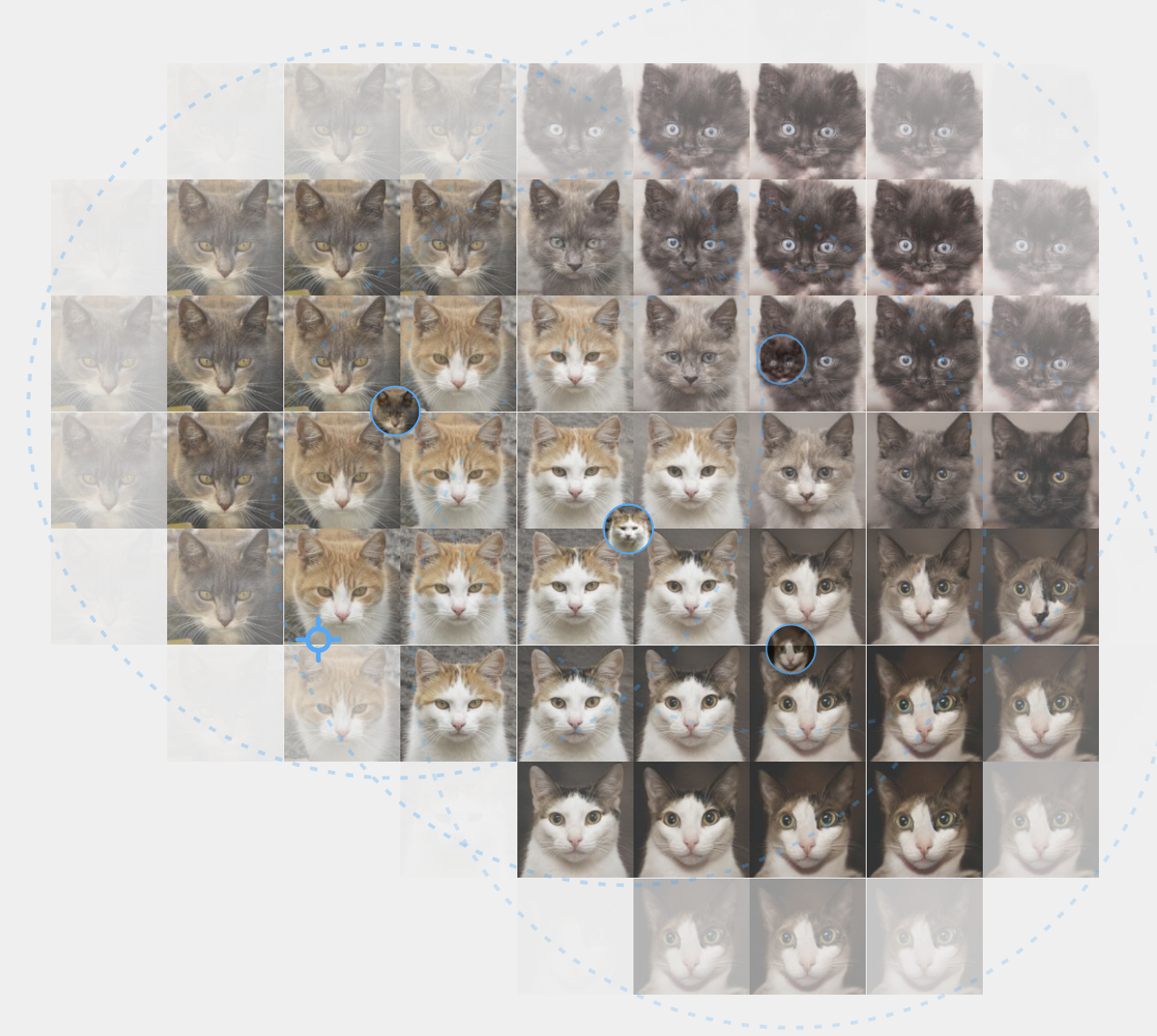}
    \caption{Local grid-mode mixing view, showing four cats get interpolated over the entire grid on the canvas.}
    \Description{A canvas filled with a regular grid of cat images interpolated from four cat examples.}
    \label{fig:local-mix-grid-cat-3}
\end{wrapfigure}

Viewing the interpolation as a 2D surface embedded in latent space, we additionally provide a \textit{grid mode} for models with sufficiently fast inference (\autoref{fig:local-mix-grid-cat-3}). Given the current neighborhood layout, the interface samples a regular grid over the canvas, computes the corresponding interpolated latent code at each location, and displays the generated output in place. This makes the full 2D interpolation surface directly visible and explorable. For example, StyleGAN2 supports this interaction comfortably in near real time on an NVIDIA RTX 4090.

We therefore position our technique primarily for \textit{free exploration}, where the relevant modes of variation are not fully known or explicitly parameterized in advance. When the desired semantics are already well specified, existing interaction techniques can provide more direct and complementary forms of control.

\begin{figure}[t]
    \centering
    \includegraphics[width=1\linewidth]{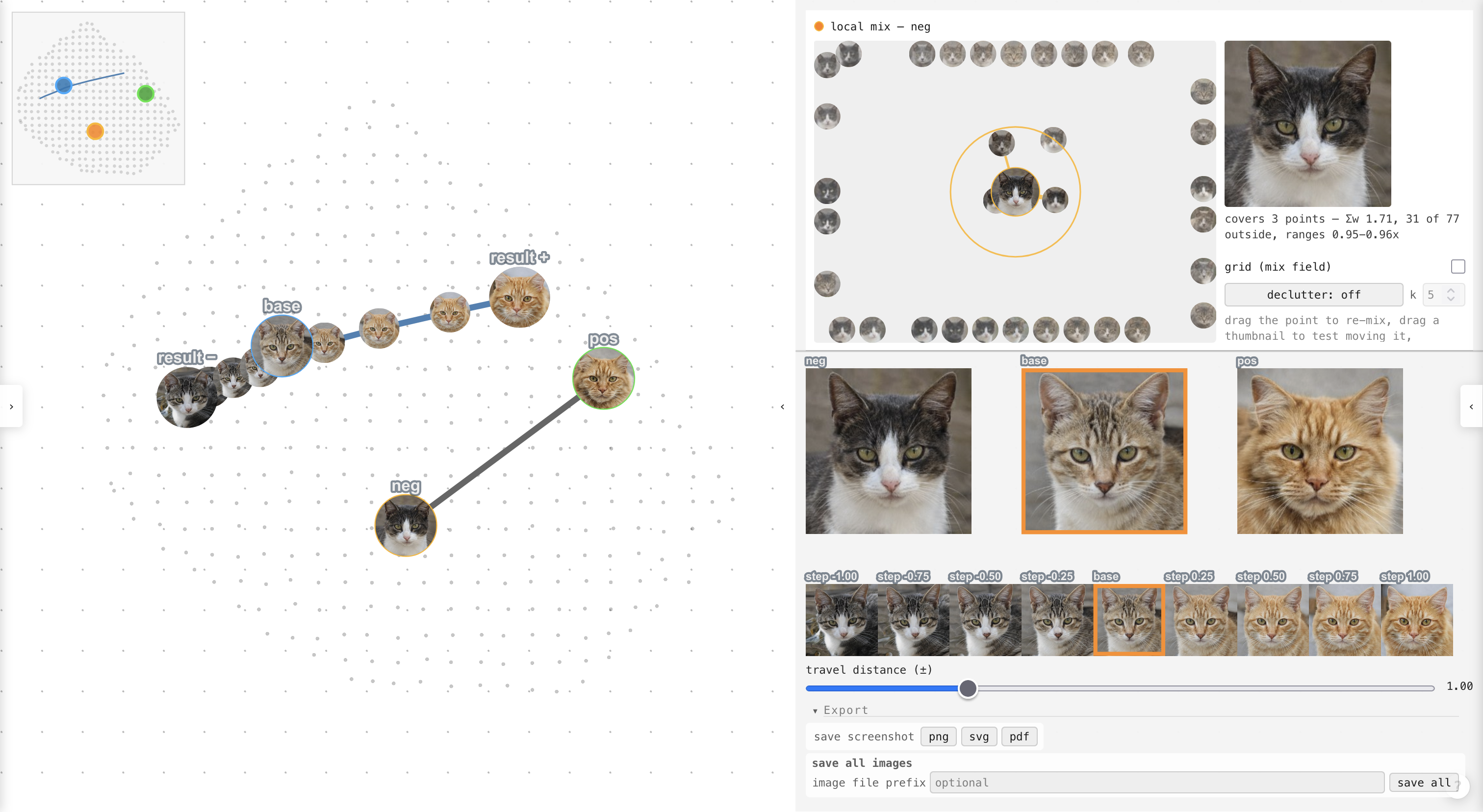}
    \caption{Screenshot of the \sysname interface. Putting these components together, \sysname combines a global navigation overview (left) for vector-arithmetic steering in latent space with a local refinement view (top right) for tuning individual components of the steering operation. The interface also includes utility controls (bottom right), such as a maximum step-size slider, image preview panel, and buttons for saving or exporting generated outputs.}
    \Description{A screenshot of the \sysname interface with a global navigation map on the left, a local refinement view at the top right, and a step-size slider, image preview, and save and export controls at the bottom right.}
    \label{fig:ui}
\end{figure}

\subsection{Linked Views and Interactions}

Putting the global and local views together, we built a prototype interface, as shown in \autoref{fig:ui} and the teaser (\autoref{fig:teaser}) that integrates global exploration, local refinement, configuration controls, and output export into a single workflow.

\begin{figure}
    \centering
    \includegraphics[width=0.8
    \linewidth]{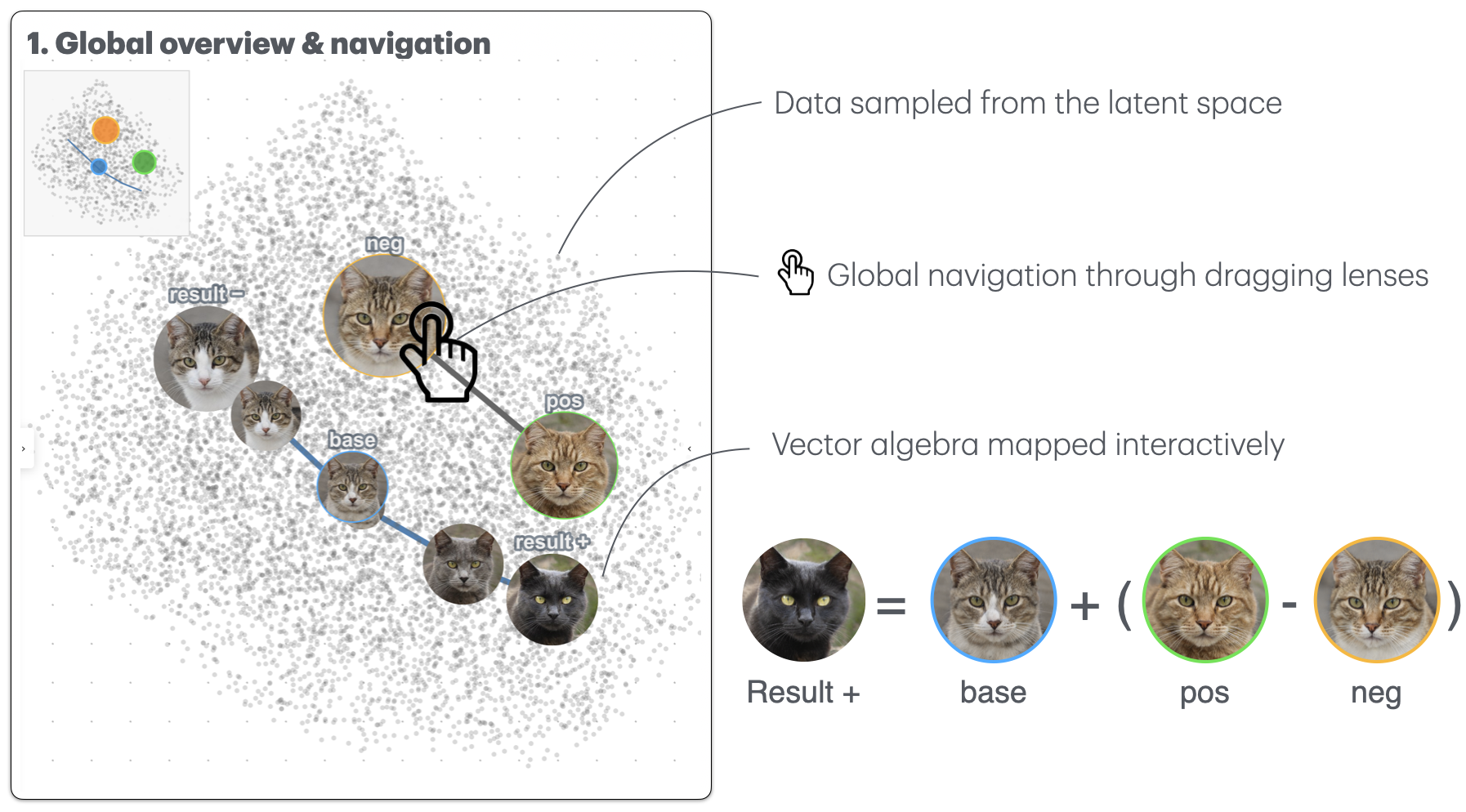}
    \caption{Global view interactions showing the result from adding the \textit{positive} direction to the base (Result+).}
    \Description{A screenshot of the global navigation view showing the projected sample map and its interaction controls.}
    \label{fig:ui-global}
\end{figure}

\subsubsection{Global View Interactions} 

\begin{figure}[h]
    \centering
    \begin{subfigure}[b]{0.4\textwidth}
        \includegraphics[width=\linewidth]{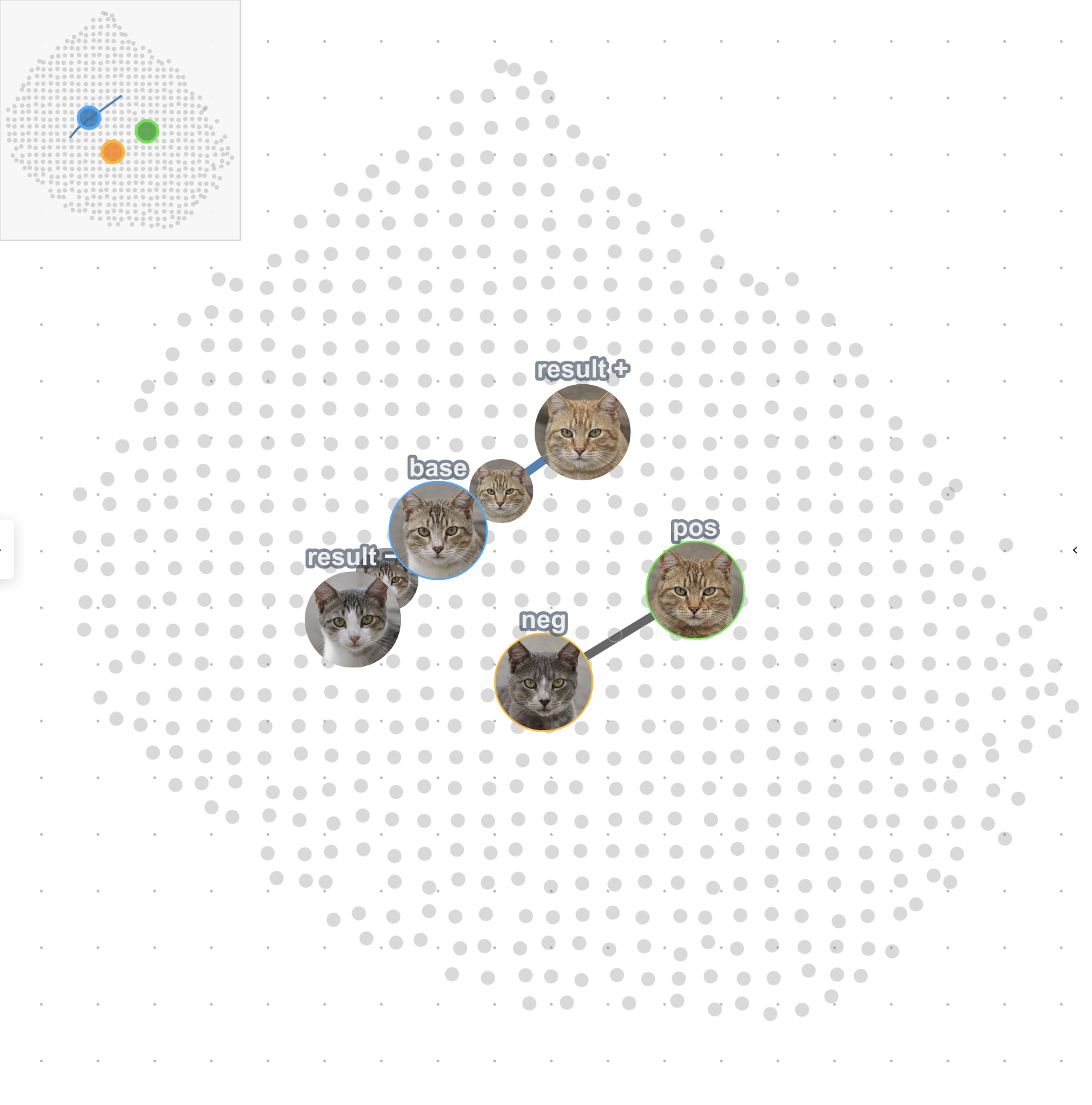}
    \end{subfigure}%
    \begin{subfigure}[b]{0.4\textwidth}
        \includegraphics[width=\linewidth]{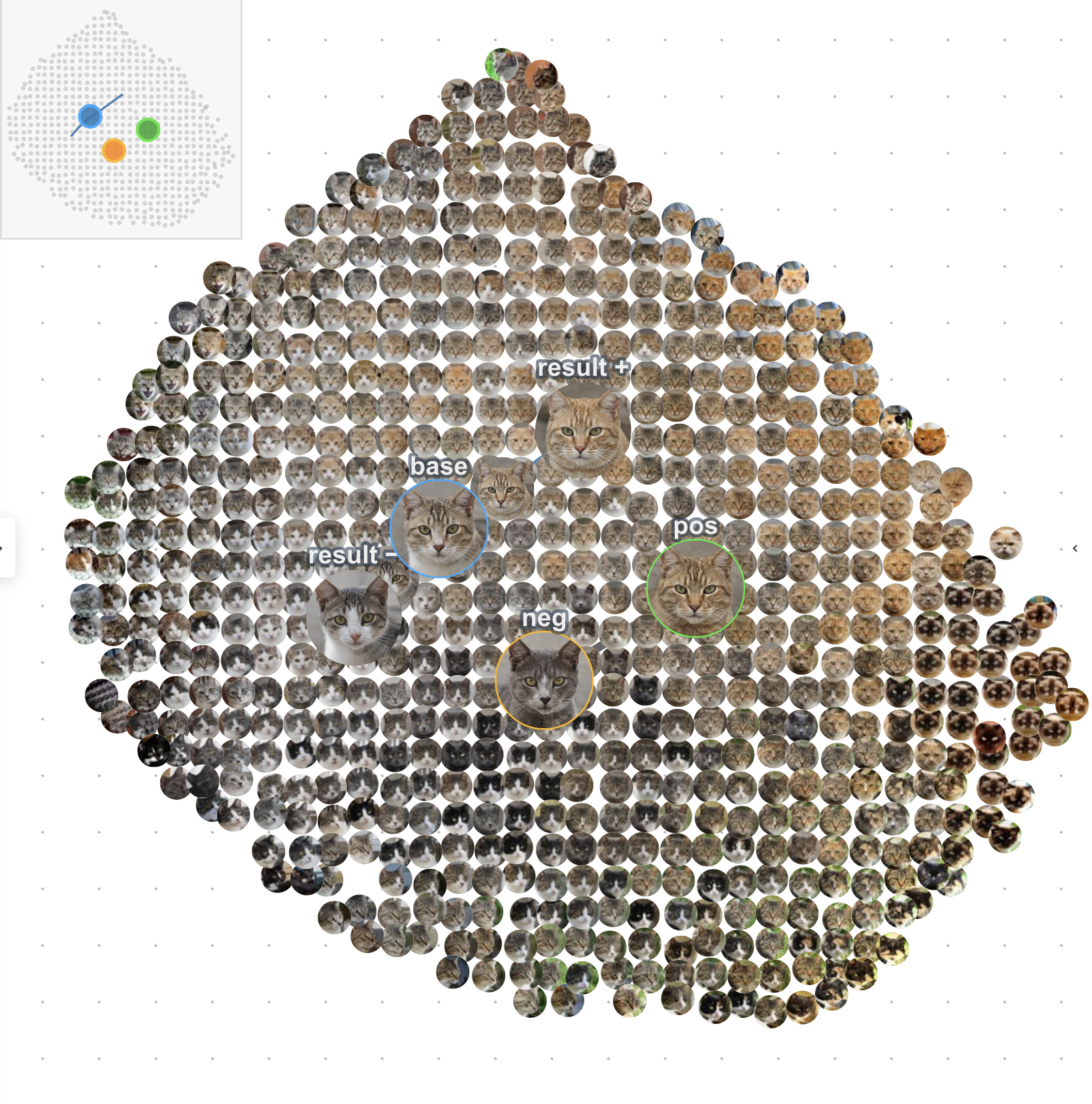}
    \end{subfigure}%
    \caption{Grid binning of samples in the global view, with (right) or without (left) image thumbnail shown}
    \Description{Two global maps with samples arranged in grid bins: the left uses marks without image thumbnails, and the right displays image thumbnails.}
    \label{fig:ui-global-bin}
\end{figure}

As described in SG1, the global view supports browsing and reasoning across the generative landscape (\autoref{fig:ui-global}). To reduce visual clutter in dense projections, \sysname provides optional grid-based binning of projected samples (\autoref{fig:ui-global-bin}). Users can also switch between simple circular marks and image thumbnails when generated outputs are images. Both marks and thumbnails can be resized using keyboard shortcuts (\texttt{Cmd + +/-}), allowing users to trade off overview density against visual detail.

\paragraph{Zooming and Panning}
The global map supports standard zooming and panning together with draggable \textit{lenses} for inspecting particular regions of the projection. Each lens represents a latent code computed as a weighted combination of the samples covered by the lens, allowing users to move continuously through regions between observed examples rather than selecting only discrete samples. The lens radius can be adjusted using the \texttt{+/-} keys while hovering over the lens, changing both the spatial extent of the neighborhood and the samples contributing to the interpolation.

\begin{figure}
    \centering
    \includegraphics[width=0.4\linewidth]{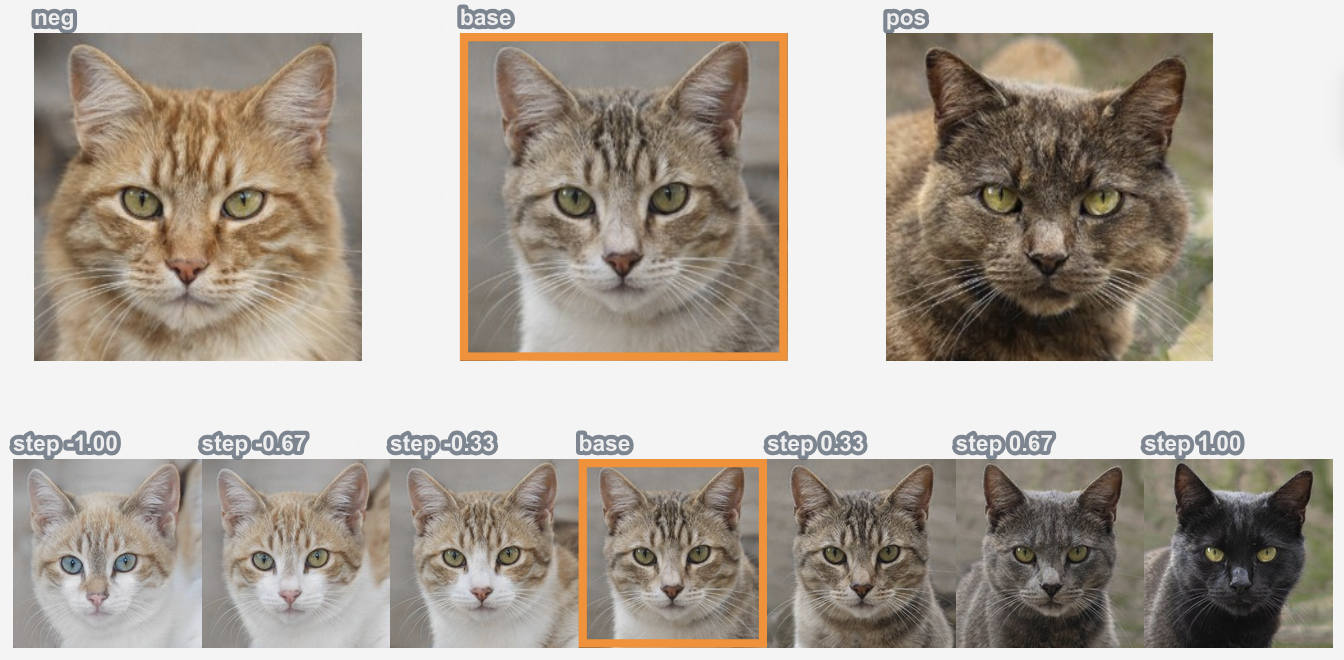}
    \caption{The interface sweeps $\alpha$ from $-1$ to $1$ over uniformly spaced steps, allowing users to inspect how the generated output changes along the resulting direction.}
    \Description{Generated outputs along a steering direction, sampled at uniformly spaced step sizes from $\alpha=-1$ to $\alpha=1$.}
    \label{fig:ui-stepsize-sweep}
\end{figure}

In steer mode, the base, positive, and negative lenses are manipulated through the same interaction. \sysname first computes the latent representation associated with each lens through interpolation and then constructs a steering trajectory using the vector arithmetic defined in \autoref{eq:algebra}. The interface sweeps $\alpha$ from $-1$ to $1$ over uniformly spaced steps, allowing users to inspect how the generated output changes along the resulting direction (\autoref{fig:ui-stepsize-sweep}).

The global view can also display multiple trajectories simultaneously (\autoref{fig:teaser}), supporting comparison of steering operations across different regions or starting points. 
Pressing \texttt{a} adds a new active trajectory, while \texttt{x} removes the most recently added trajectory.

\subsubsection{Local View Interactions}

\begin{figure}
    \centering
    \begin{subfigure}[b]{0.5\textwidth}
    \includegraphics[width=\linewidth]{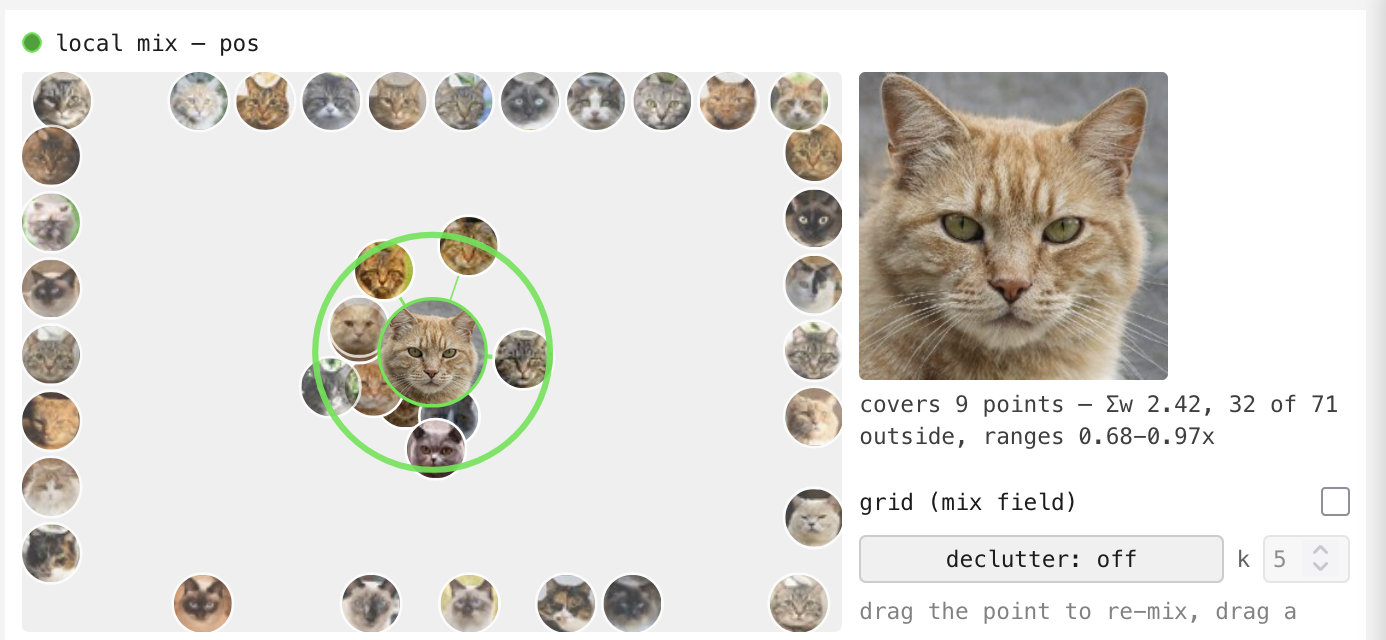}
    \end{subfigure}%
    \begin{subfigure}[b]{0.5\textwidth}
    \includegraphics[width=\linewidth]{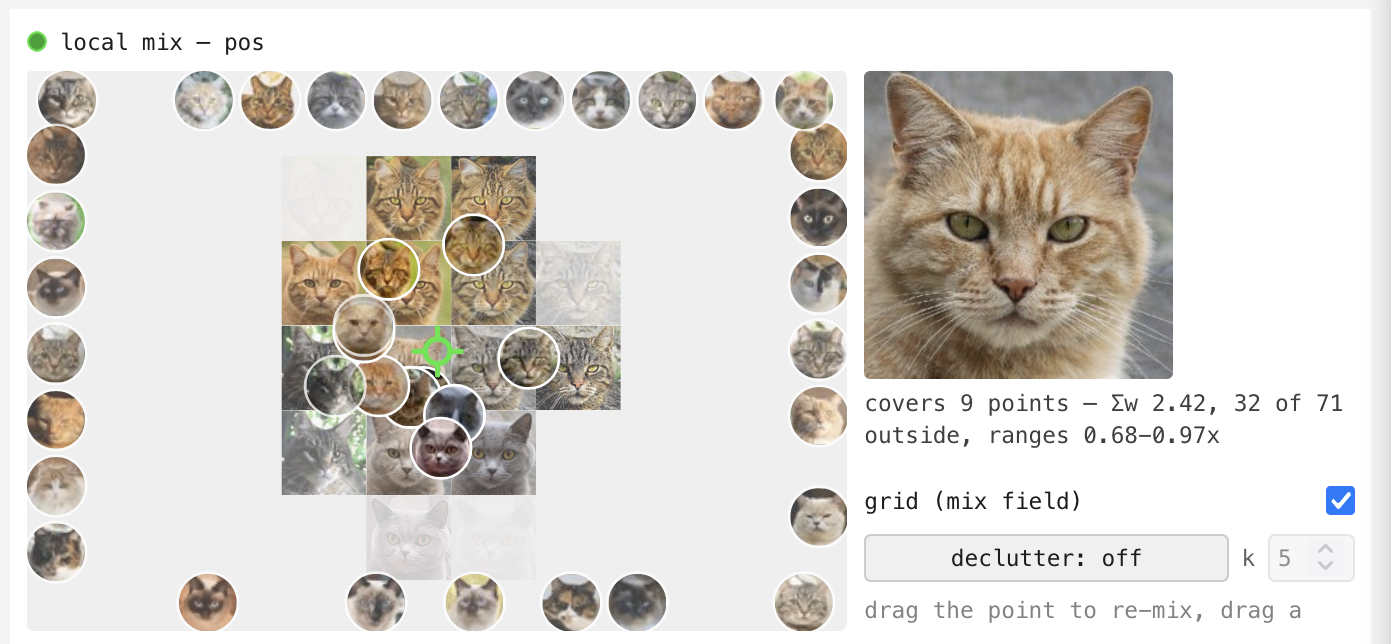}
    \end{subfigure}%
    \caption{Local refinement view in (left) lens mode and (right) grid mode.}
    \Description{Two local refinement views side by side: a central lens surrounded by neighboring anchors on the left, and a grid of interpolated outputs on the right.}
    \label{fig:ui-local-small}
\end{figure}

\begin{figure}
    \centering
    \includegraphics[width=1.0\linewidth]{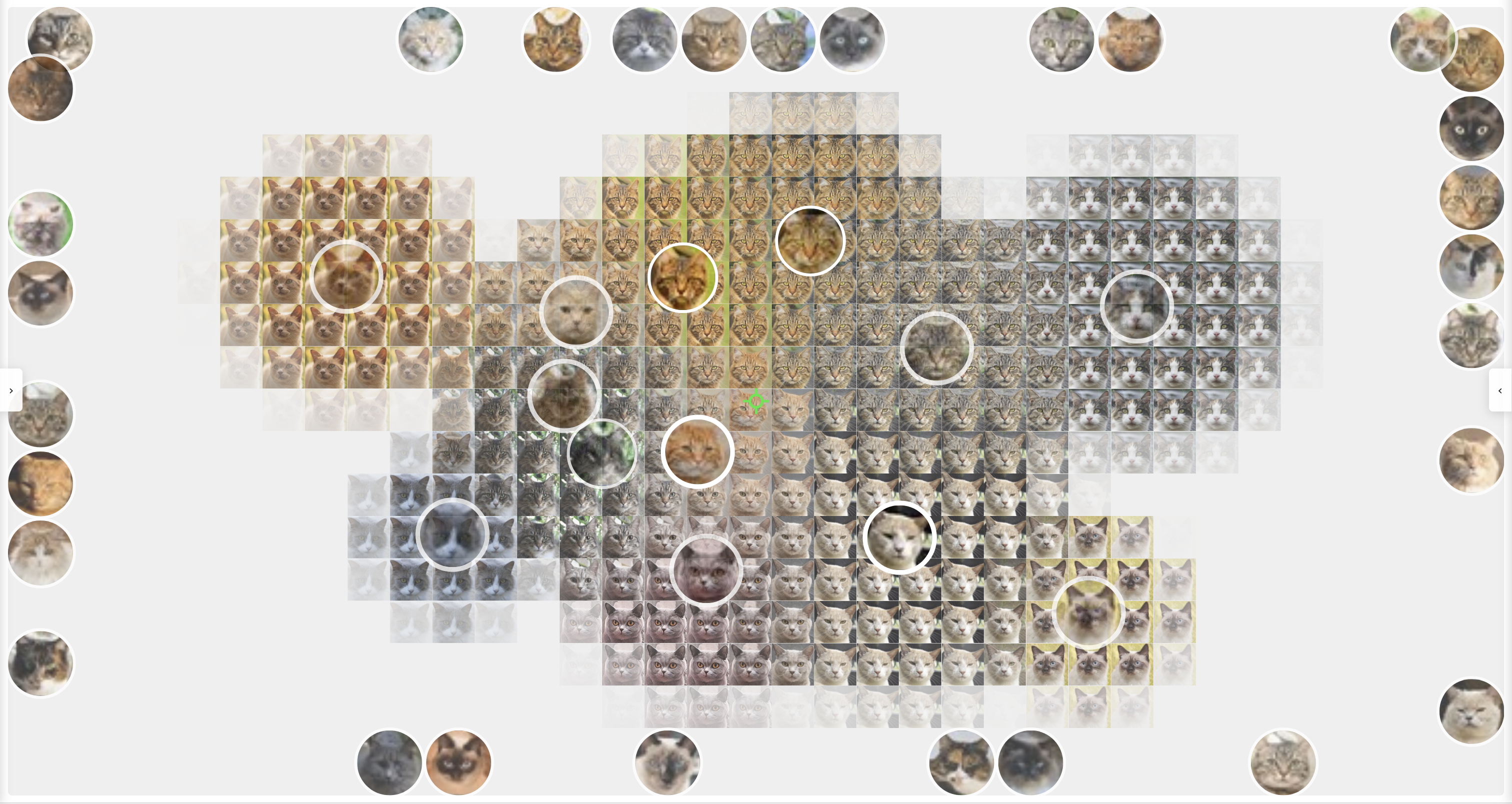}
    \caption{Local refinement in grid mode with the view maximized.}
    \Description{A maximized local refinement canvas displaying a grid of interpolated outputs and the anchors defining their mixtures.}
    \label{fig:ui-local-grid-max}
\end{figure}

The local view provides finer-grained control over the interpolation neighborhood around a selected point (\autoref{fig:ui-local-small}). A lens is fixed at the center of the view and represents the current interpolated latent code. Dragging the central lens moves this code relative to the surrounding anchors; once the interaction ends, the view is recentered around the updated position.

Users can also directly manipulate the neighboring anchor points. Dragging an anchor changes the local layout and, consequently, the interpolation weights contributing to the central lens. Anchor thumbnails can be resized using \texttt{Cmd + +/-}. The central lens itself can be resized using \texttt{+/-}, which changes the cutoff radius $R$ of the distance-to-weight mapping and therefore the spatial scope of the interpolation. Double-clicking an anchor restores it to its original local position, while double-clicking the central lens pushes all currently active anchors outside the cutoff radius, effectively clearing the current interpolation neighborhood.

To reduce visual clutter, neighbors outside the current local region are displayed along the boundary of the local mixer rather than removed entirely. Their underlying positions remain unchanged. As the user drags the central lens toward another part of the neighborhood, relevant anchors are progressively brought into the active region, while more distant points recede toward the periphery. This preserves access to nearby alternatives without overwhelming the central interaction area.

\paragraph{Grid mode.}
The local view additionally supports a grid mode (\autoref{fig:ui-local-small} right, and \autoref{fig:ui-local-grid-max}) in which each grid location on the canvas is assigned an interpolated latent code using the same weighting scheme. This turns the local mixer into a direct visualization of the 2D interpolation surface. Similar anchor-manipulation interactions remain available, although here each anchor's effective support replaces the notion of a single central-lens radius. The grid resolution can be adjusted using \texttt{Cmd + +/-} when the cursor is not hovering over an anchor. Because every grid location may require a separate model evaluation, this mode can be computationally expensive depending on the model architecture, hardware, and grid resolution. We were able to run StyleGAN2 interactively in this mode on an NVIDIA RTX 4090.

\begin{figure}
    \centering
    \includegraphics[width=0.75\linewidth]{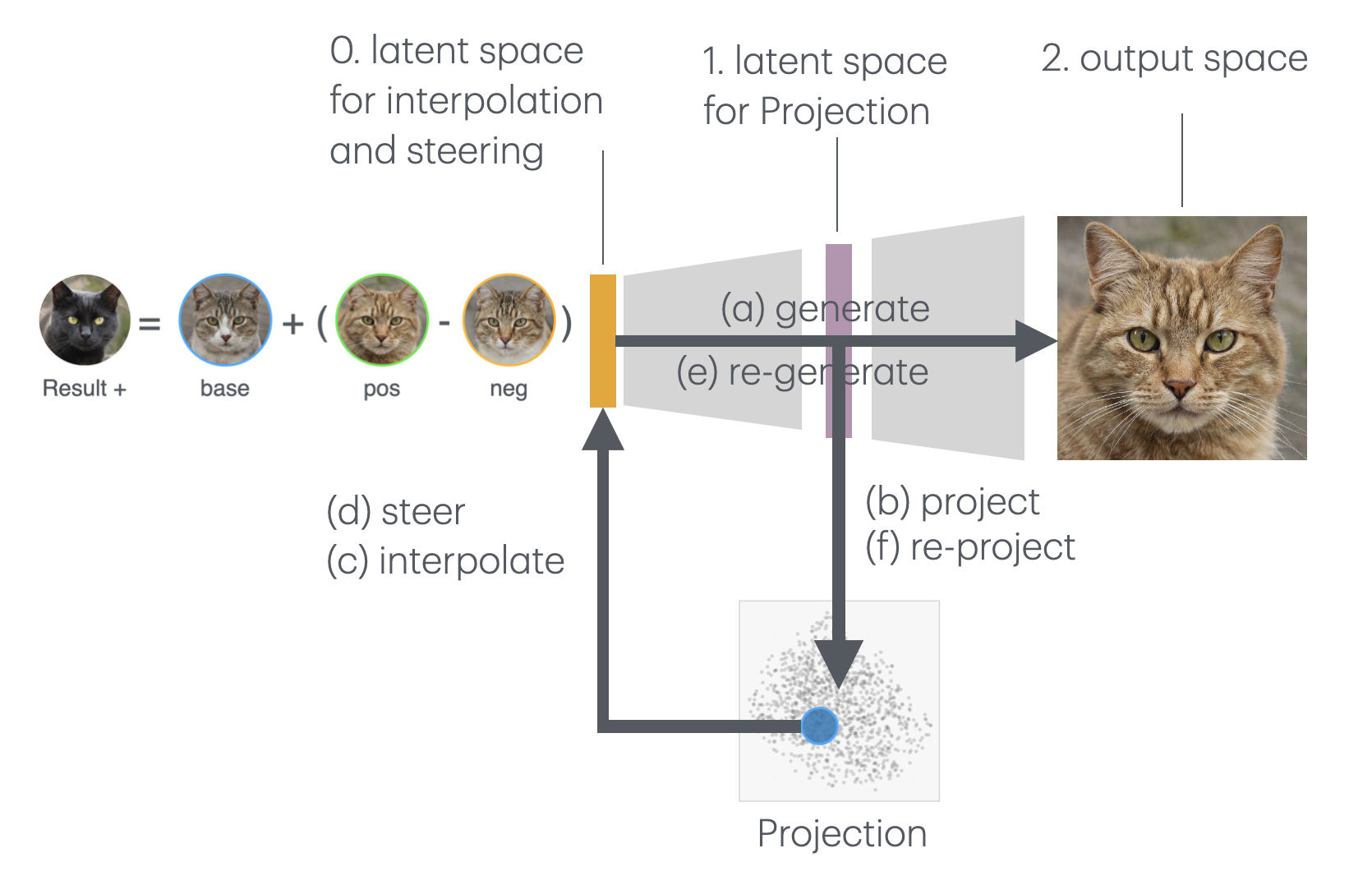}
    \caption{Data flow diagram of the \sysname system. Mapping consistency is maintained through parametric dimensionality reduction methods, such as Parametric UMAP \cite{sainburg2021parametric}, which can project newly generated points into the same embedding rather than being limited to the original training samples.}
    \Description{A data flow diagram connecting the latent representation used for interpolation and steering, the representation projected into the two-dimensional overview, and the generated output space.}
    \label{fig:data-flow}
\end{figure}

\subsubsection{Global-local Consistency}

The global and local views operate on the same underlying latent representations, allowing changes made in one view to propagate to the other. In the global view, the latent codes produced by interpolation or steering operations are projected back into the overview using the learned parametric dimensionality-reduction mapping. 

\autoref{fig:data-flow} summarizes the data flow through \sysname. The system operates across three related spaces: (0) a latent representation suitable for interpolation and steering; (1) a representation used to construct the dimensionality-reduction overview; and (2) the generated output space. These spaces need not coincide. In the StyleGAN case, for example, we perform interpolation and steering in the style representation, while constructing the overview from later feature maps, which exhibit richer structure under dimensionality reduction than the comparatively uniform style space.

The core idea of \sysname is to let users interact with a 2D projection while carrying out interpolation and steering in the corresponding high-dimensional representation. User interactions in the overview therefore define operations in space (0), while their results are projected back into space (1) for visualization and decoded into space (2) for inspection. Mapping consistency is maintained through parametric dimensionality-reduction methods such as Parametric UMAP~\cite{sainburg2021parametric}, which can embed newly generated and interpolated points into the same learned overview rather than being restricted to the original samples.

Likewise, after a user refines an interpolated point in the local view, the resulting latent representation can be mapped back into the global overview to show how the refinement changes its position within the broader generative landscape.

A local refinement changes the arrangement of points in the local mixer but should preserve their original positions in the global map to maintain the user's mental map. We think of this as navigating with an old map after the local surroundings have changed: the global view preserves the original reference structure, while the updated local configuration is revealed when the user revisits that neighborhood. This separation provides a stable global frame of reference while allowing users to reorganize local neighborhoods for more effective interpolation. 
Internally, both views therefore manipulate a shared representation of each point, while maintaining view-specific spatial coordinates. Each sampled point has a fixed global coordinate derived from the dimensionality-reduction embedding, together with a local coordinate that may be modified through direct manipulation in the local mixer. This distinction allows users to reorganize a neighborhood for interpolation without altering the underlying global map.

This separation can otherwise create a consistency problem. For example, if a user drags a neighboring point far away during local refinement, they may later return expecting to find that point near its original location. Because \sysname retains the original coordinates, we display such points as peripheral ``ghost'' anchors in the local view rather than removing them entirely. Users can therefore recover or reposition them while retaining a stable correspondence with the original global map. In future work, we also envision allowing a single sample to appear at multiple local positions, supporting multiple locally configured neighborhoods while preserving a shared global identity.

\subsubsection{Other Features}

In addition to the core navigation interactions, \sysname includes several visual and interaction features intended to improve usability during extended exploration.

\paragraph{Minimap.}
A minimap in the upper-right corner of the global overview provides context for the current viewport and helps users maintain orientation while zooming and panning across large latent space maps. Users can also drag lenses directly within the minimap.

\paragraph{Progressive and minibatched decoding.}
Computationally intensive interactions may require regenerating many outputs simultaneously. For example, moving the positive lens in steer mode can update every sample along the manipulation trajectory. \sysname therefore supports progressive and minibatched decoding so that partial results can be displayed before the complete trajectory has finished rendering. For diffusion models, the interface can additionally expose intermediate denoising results while the final outputs are being computed. These mechanisms provide earlier visual feedback during otherwise high-latency interactions.

\paragraph{Customizable lens labels.}
Lens labels can be edited directly in the interface, allowing users to annotate examples, concepts, or steering directions during exploration. These annotations can also be retained in exported views, supporting documentation and communication of the exploration process.






\section{Evaluation}\label{sect:eval}\label{sect:use-cases}
We evaluate \sysname through two use cases involving generative models of varying complexity, together with a formative study examining \sysname 's application to an engineering design task involving rotational kirigami systems.
The three evaluation cases progress in increasing complexity and application realism. 
StyleGAN serves as a proof of concept and, because of its fast inference, as a testbed for runtime interaction enhancements such as real-time neighbor sampling. 
The diffusion-model case demonstrates the generalizability of the \sysname framework to a more complex generative architecture.
Finally, the kirigami case evaluates \sysname in a real-world structural engineering design setting. The polygonal design space in this use case also provides a useful testbed for examining \sysname under conditions where the underlying degrees of freedom, modes and extent of variation, and global topology of the design space are explicitly known.

\subsection{Case 1: StyleGAN2 faces and cats}

\subsubsection{The generative model and data representation}
We sampled $20,000$ points from the input space of StyleGAN2 models trained separately on human faces and cat pictures.
Importantly, the representation used to build the global map need not be the same representation used for semantic interpolation. In the StyleGAN2 case, the overview map is constructed from sampled activations after the model's \texttt{b8} block, yielding $512 \times 8 \times 8$-dimensional feature representations that are projected into 2D or 3D using UMAP. In contrast, the interpolation algebra is performed in StyleGAN2's 512-dimensional style space, obtained after its mapping network. This separation allows the overview representation to be chosen for effective visualization while the interpolation representation is chosen for meaningful and controllable semantic edits.


\subsubsection{Use case walkthrough}
\begin{figure}
    \centering
    \includegraphics[width=0.6\linewidth]{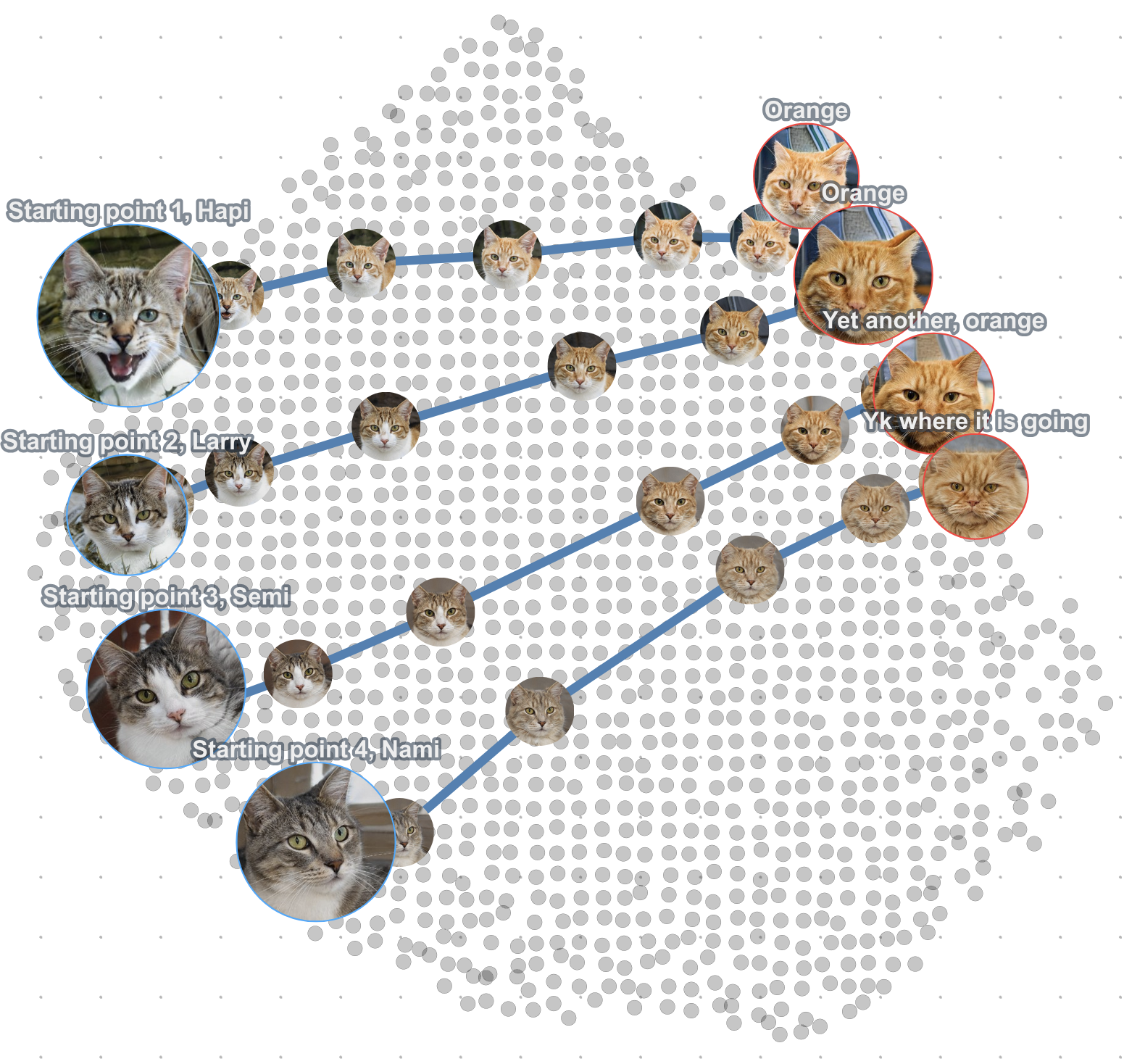}
    \caption{Linearly and uniformly interpolating two end points in the latent space via a non-linear dimensionality reduction plot in \sysname.}
    \Description{A global dimensionality-reduction map showing interpolation paths between cat examples, including transitions toward orange cats.}
    \label{fig:orange-cat-1}
\end{figure}
We first validate the interpolation capacity of the interface by using it to construct transitions between arbitrary pairs of data points. 
Using cat images as an example, we trace multiple paths from different starting points toward regions associated with orange cats 
in the global overview and inspect the corresponding generated transitions. 
As shown in the teaser and \autoref{fig:orange-cat-1}, the interface produces smooth and expected changes (turning cats orange) along these paths, demonstrating that a simple map-based interaction can effectively control interpolation in the underlying generative space.
Importantly, the space used for interpolation and steering does not need to be the same space used to construct the global map. The overview may be built from a dimensionality-reduced representation of one model layer, while interpolation is performed in another latent or feature space better suited for generation. The map therefore serves primarily as a spatial interface for selecting and visualizing trajectories, rather than as the space in which the interpolation itself is computed. Because the overview is produced by a nonlinear projection such as UMAP, a straight interpolation path in the underlying latent space may also appear curved or distorted when mapped back into the overview.

\paragraph{\sysname as a dimensionality reduction linearity checker}
Here, we also observed that \sysname can incidentally serve as a probe of local linearity in dimensionality-reduction plots. The negative-to-positive direction acts as a ``straight ruler'' in the high-dimensional latent space: by applying the same linear direction around different base points and examining its projection, we can observe how a straight latent space trajectory is locally distorted by a nonlinear dimensionality-reduction method. 
Because linearity in a high-dimensional space must be considered across multiple directions, interactivity provides a practical way to probe this distortion one direction at a time, revealing how local curvature in the projection varies across both position and direction.

\begin{figure}[h]
    \centering
    \includegraphics[width=1\linewidth]{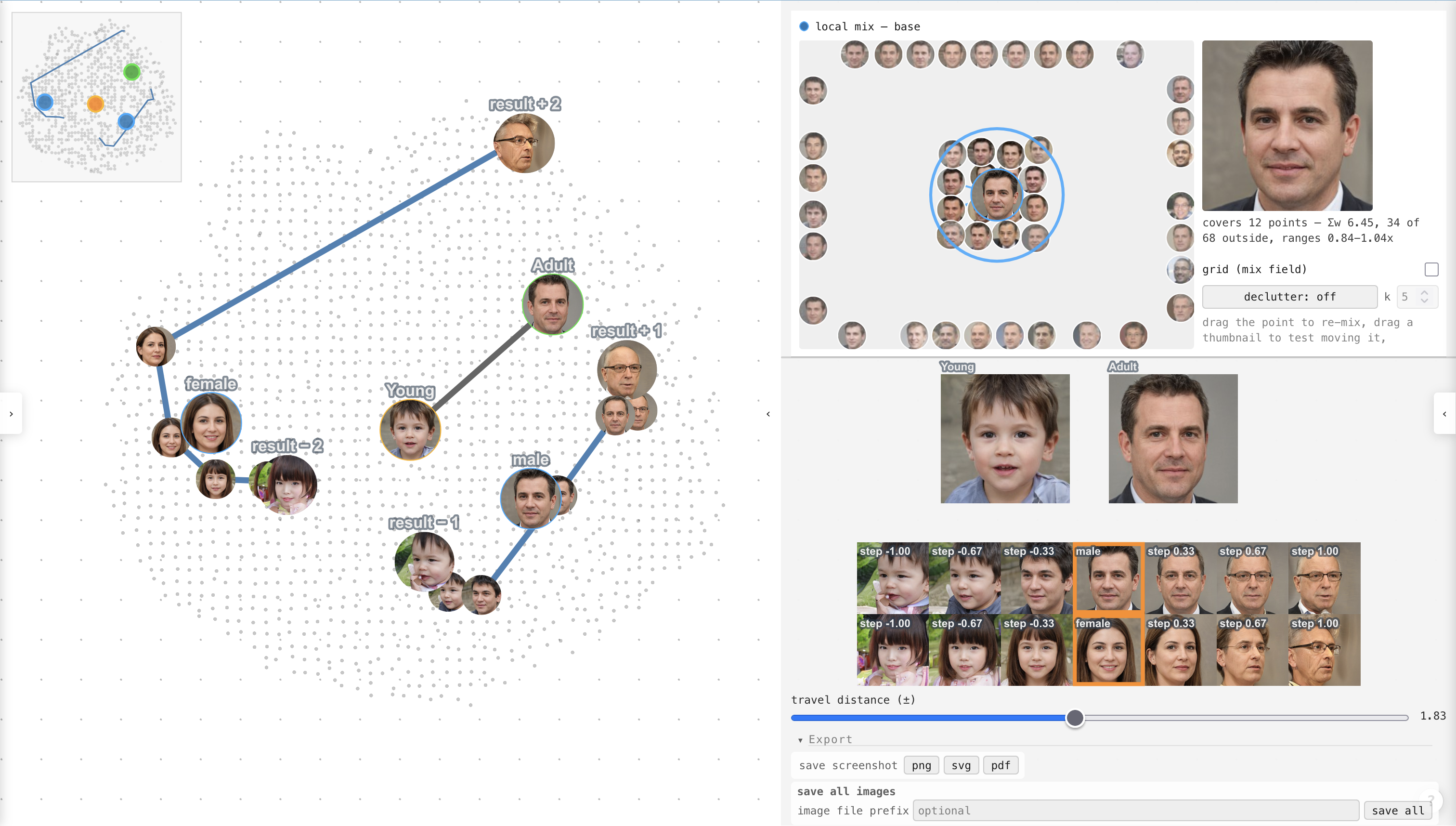}
        \begin{subfigure}[b]{0.142\textwidth}
        \includegraphics[width=\linewidth]{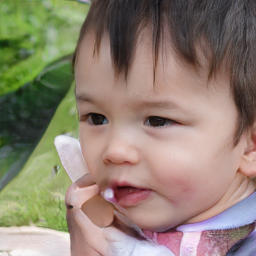}
        \end{subfigure}%
        \begin{subfigure}[b]{0.142\textwidth}
        \includegraphics[width=\linewidth]{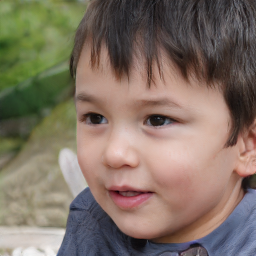}
        \end{subfigure}%
        \begin{subfigure}[b]{0.142\textwidth}
        \includegraphics[width=\linewidth]{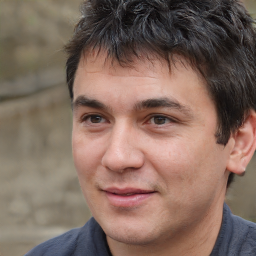}
        \end{subfigure}%
        \begin{subfigure}[b]{0.142\textwidth}
        \includegraphics[width=\linewidth]{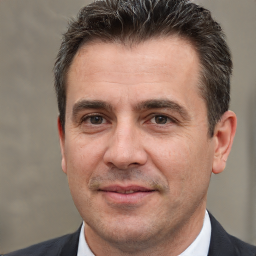}
        \end{subfigure}%
        \begin{subfigure}[b]{0.142\textwidth}
        \includegraphics[width=\linewidth]{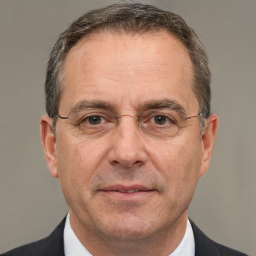}
        \end{subfigure}%
        \begin{subfigure}[b]{0.142\textwidth}
        \includegraphics[width=\linewidth]{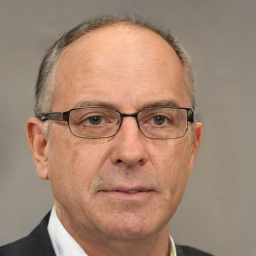}
        \end{subfigure}%
        \begin{subfigure}[b]{0.142\textwidth}
        \includegraphics[width=\linewidth]{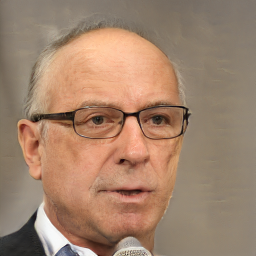}
        \end{subfigure}%

        \begin{subfigure}[b]{0.142\textwidth}
        \includegraphics[width=\linewidth]{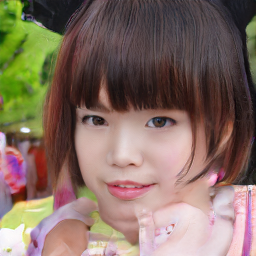}
        \end{subfigure}%
        \begin{subfigure}[b]{0.142\textwidth}
        \includegraphics[width=\linewidth]{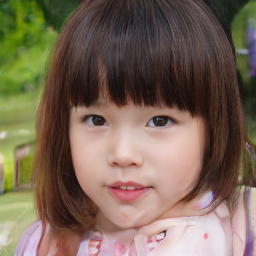}
        \end{subfigure}%
        \begin{subfigure}[b]{0.142\textwidth}
        \includegraphics[width=\linewidth]{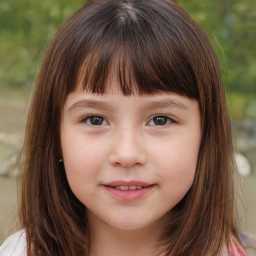}
        \end{subfigure}%
        \begin{subfigure}[b]{0.142\textwidth}
        \includegraphics[width=\linewidth]{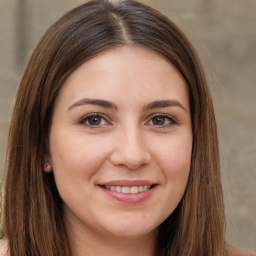}
        \end{subfigure}%
        \begin{subfigure}[b]{0.142\textwidth}
        \includegraphics[width=\linewidth]{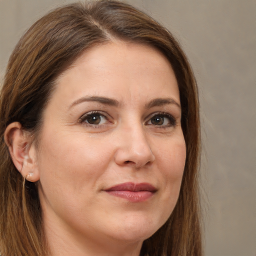}
        \end{subfigure}%
        \begin{subfigure}[b]{0.142\textwidth}
        \includegraphics[width=\linewidth]{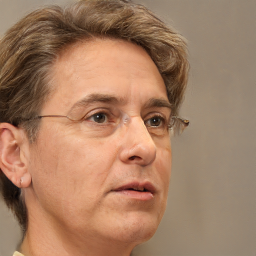}
        \end{subfigure}%
    
    \begin{subfigure}[b]{0.142\textwidth}
        \fcolorbox{red}{white}{\includegraphics[width=\linewidth]{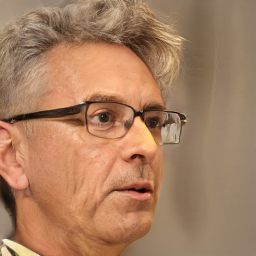}}
    \end{subfigure}%
    \caption{In the face dataset, the age direction becomes entangled with gender near the extreme end of the female-face trajectory (boxed in red).}
    \Description{An interface screenshot above two rows of seven generated faces along age-steering trajectories. A red box highlights the final face in the second row, where the female-face trajectory exhibits an unintended gender change.}
    \label{fig:face-entangled}
\end{figure}

Next, we examine the manipulation capability of the global view. Using human faces as an example, suppose we want to identify a global direction for age. We first define an aging direction from contrastive examples by browsing the landscape to locate regions containing younger and older faces. Placing the positive lens over older faces and the negative lens over younger faces defines a direction corresponding to increasing age. We then test how this direction transfers across different base points by moving the base lens through the map. \autoref{fig:face-entangled} shows two example base points, one male and one female, for which the same age direction produces meaningful edits.


\begin{figure}[h]
    \centering
    \includegraphics[width=1\linewidth]{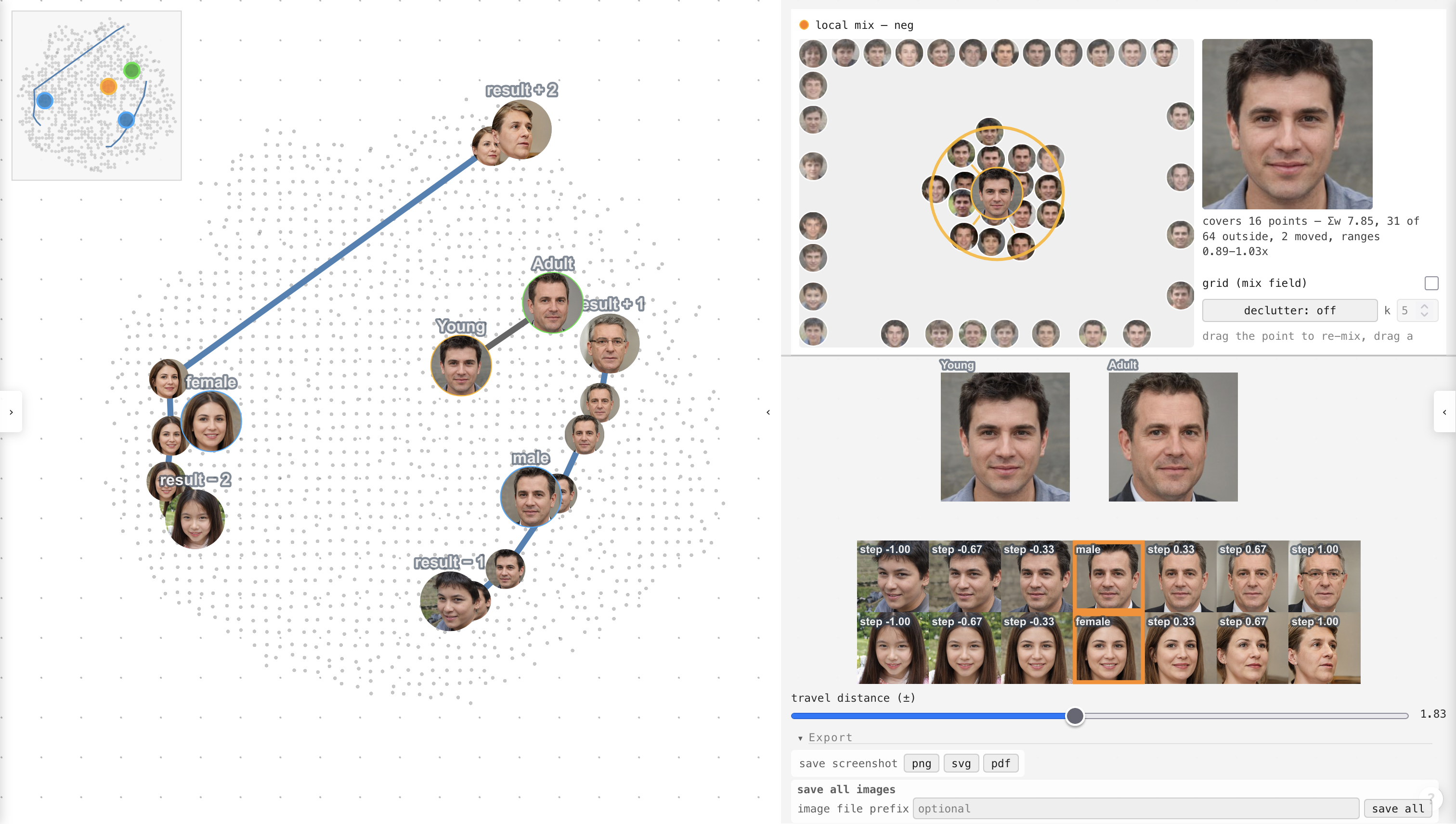}
        \begin{subfigure}[b]{0.142\textwidth}
        \includegraphics[width=\linewidth]{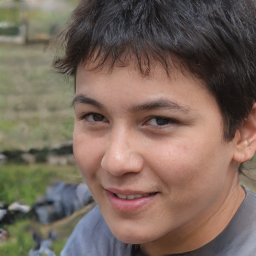}
        \end{subfigure}%
        \begin{subfigure}[b]{0.142\textwidth}
        \includegraphics[width=\linewidth]{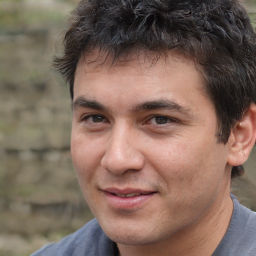}
        \end{subfigure}%
        \begin{subfigure}[b]{0.142\textwidth}
        \includegraphics[width=\linewidth]{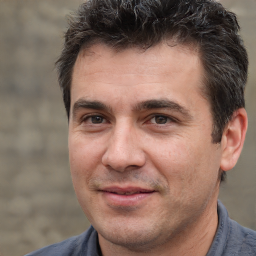}
        \end{subfigure}%
        \begin{subfigure}[b]{0.142\textwidth}
        \includegraphics[width=\linewidth]{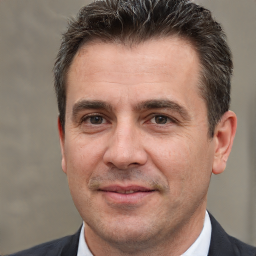}
        \end{subfigure}%
        \begin{subfigure}[b]{0.142\textwidth}
        \includegraphics[width=\linewidth]{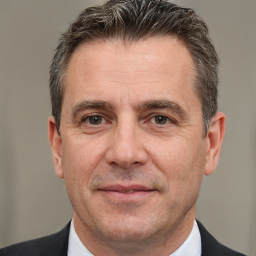}
        \end{subfigure}%
        \begin{subfigure}[b]{0.142\textwidth}
        \includegraphics[width=\linewidth]{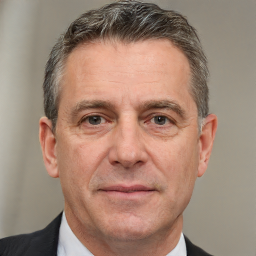}
        \end{subfigure}%
        \begin{subfigure}[b]{0.142\textwidth}
        \includegraphics[width=\linewidth]{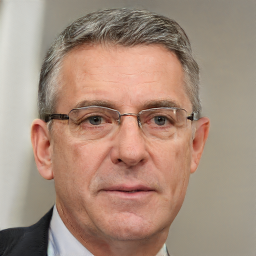}
        \end{subfigure}%

        \begin{subfigure}[b]{0.142\textwidth}
        \includegraphics[width=\linewidth]{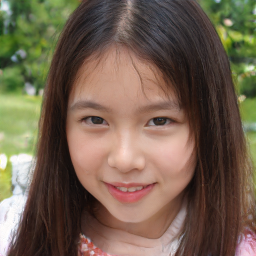}
        \end{subfigure}%
        \begin{subfigure}[b]{0.142\textwidth}
        \includegraphics[width=\linewidth]{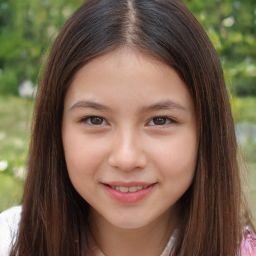}
        \end{subfigure}%
        \begin{subfigure}[b]{0.142\textwidth}
        \includegraphics[width=\linewidth]{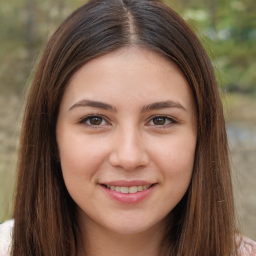}
        \end{subfigure}%
        \begin{subfigure}[b]{0.142\textwidth}
        \includegraphics[width=\linewidth]{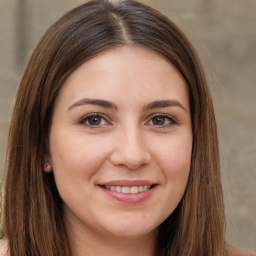}
        \end{subfigure}%
        \begin{subfigure}[b]{0.142\textwidth}
        \includegraphics[width=\linewidth]{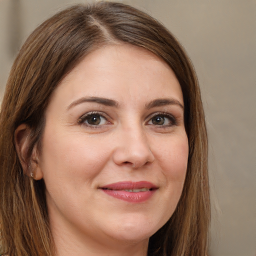}
        \end{subfigure}%
        \begin{subfigure}[b]{0.142\textwidth}
        \includegraphics[width=\linewidth]{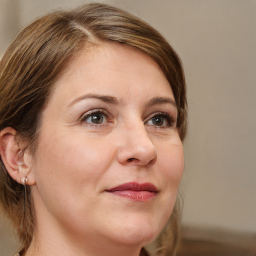}
        \end{subfigure}%
    
    \begin{subfigure}[b]{0.142\textwidth}
        \fcolorbox{red}{white}{\includegraphics[width=\linewidth]{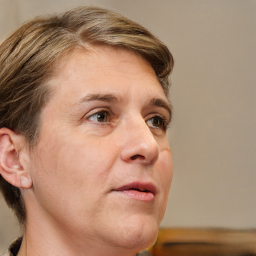}}
    \end{subfigure}%
    \caption{face dataset, the entanglement of age and gender is resolved (at the very end of the female face (boxed in red), compare to \autoref{fig:face-age-gender-entangled}) by finetuning the negative example in the local mix view.  }
    \Description{An interface screenshot above two rows of seven generated faces after local refinement of the negative example. A red box highlights the final face in the second row, which retains the intended female appearance while showing increased age.}
    \label{fig:face-disentangled}
\end{figure}

Finally, we demonstrate the refinement step in the local mixer view. Starting from the age direction defined in the global view, we adjust the neighboring examples contributing to the negative lens so that the final edited female face remains female while still exhibiting the intended increase in age (compare \autoref{fig:face-entangled} and \autoref{fig:face-disentangled}). By interactively repositioning these examples in the local mixer, we can quickly reduce the unwanted gender entanglement while preserving the desired age transformation. This illustrates how local refinement can correct undesirable correlations in a globally defined steering direction without requiring the user to redefine the manipulation from scratch.



\subsection{Case 2: Generative Chair Design using Stable Diffusion}

\subsubsection{Generative Model and Dataset generation}
We used Stable Diffusion XL \cite{Podell2023SDXLIL} (SDXL-1.0-base) to sample $4{,}000$ points using prompts describing six chair styles (\texttt{eames\_lounge}, \texttt{panton}, \texttt{judd\_84}, \texttt{lc2}, \texttt{lc4}, and \texttt{wassily}), each embedded in a shared prompt template that standardizes lighting, camera angle, and scene setup. For example, the style description for \texttt{judd\_84} is ``Judd Chair 84, strict rectilinear wooden chair assembled from flat orthogonal boards,'' which is inserted into a template beginning with ``\{style\_description\}, one chair only, one complete functional chair \ldots{} three-quarter front view, camera at seat height, plain light gray studio background, neutral material, soft studio lighting, product photograph.'' We additionally use a negative prompt to suppress common artifacts and unwanted content, including ``multiple chairs, two chairs, duplicate chair, chair set, stacked chairs, \ldots{} person, text, logo, watermark, cropped chair, hidden chair.''

We use 20 denoising steps with a guidance scale of $7.5$ and the \texttt{EulerDiscreteScheduler}, which uses a deterministic first-order Euler update without introducing additional stochastic noise during denoising. We keep the scheduler and randomization settings fixed across runs to improve reproducibility and consistency when comparing interpolated samples in \sysname. Generating the $4{,}000$ samples required approximately three hours. Compared with StyleGAN, the main practical limitation of this use case is therefore the substantially higher inference cost of diffusion models.

\subsubsection{Interpolation}
Interpolation is performed on SDXL's fully denoised latent representations. Given a set of latent codes, we first compute their weighted average according to the interpolation weights defined by the interface. Directly decoding this averaged latent tends to produce overly smooth or blurred results: averaging spatial latent features attenuates their variance and can mix incompatible structures across the interpolants. We therefore re-noise the interpolated latent to a fixed diffusion timestep and run the denoising process again. This allows the diffusion model to reconstruct a coherent sample around the interpolated representation while preserving the continuous control provided by the latent space interpolation.

\subsubsection{Observations}
\begin{figure}
    \centering
    \includegraphics[width=1\linewidth]{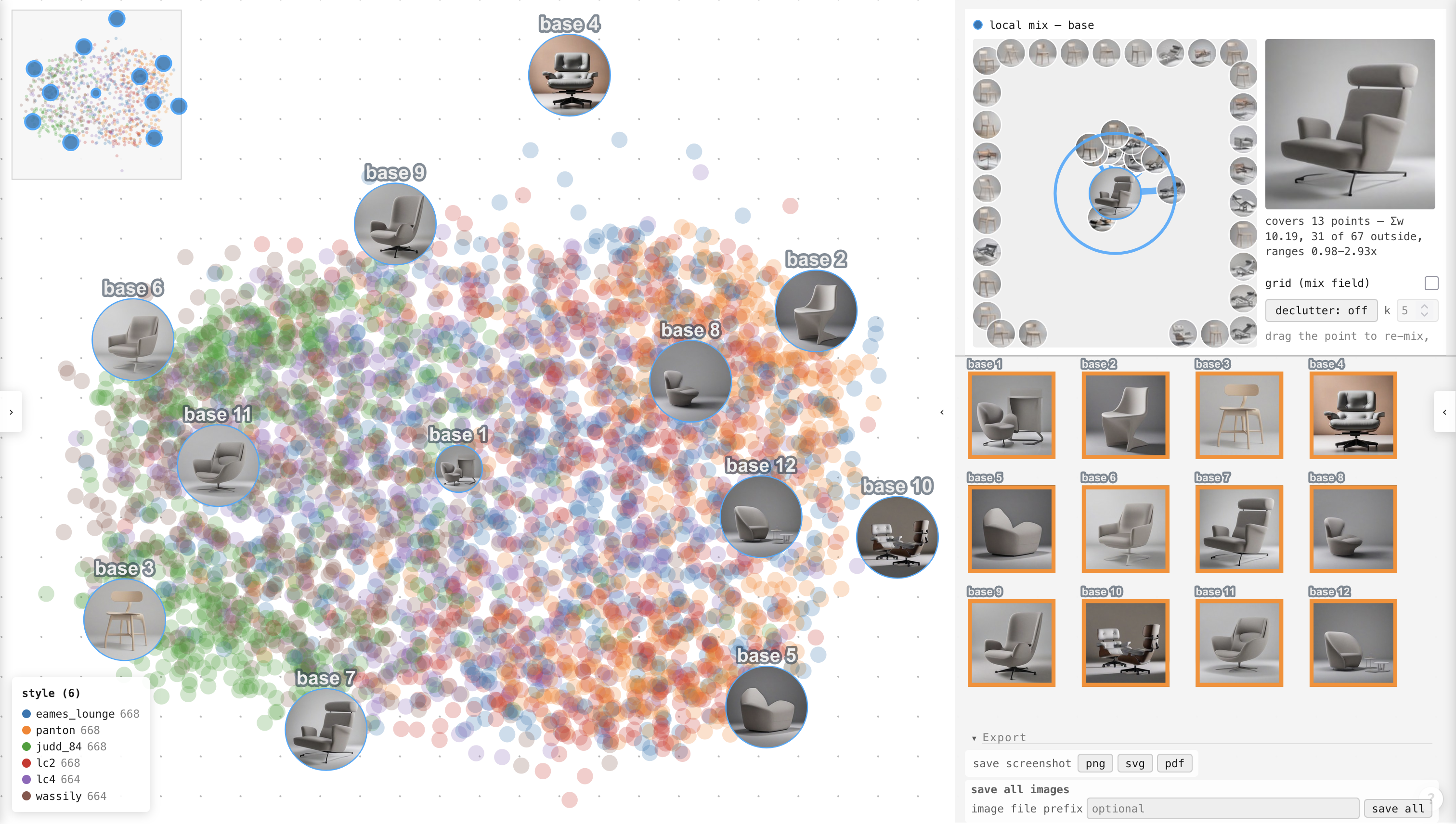}
    \caption{\sysname applied to a diffusion model for exploring chair styles.}
    \Description{A screenshot of \sysname displaying diffusion-generated chair examples for exploring variations in chair style.}
    \label{fig:chair-1}
\end{figure}

\begin{figure}
    \centering
    \includegraphics[width=1\linewidth]{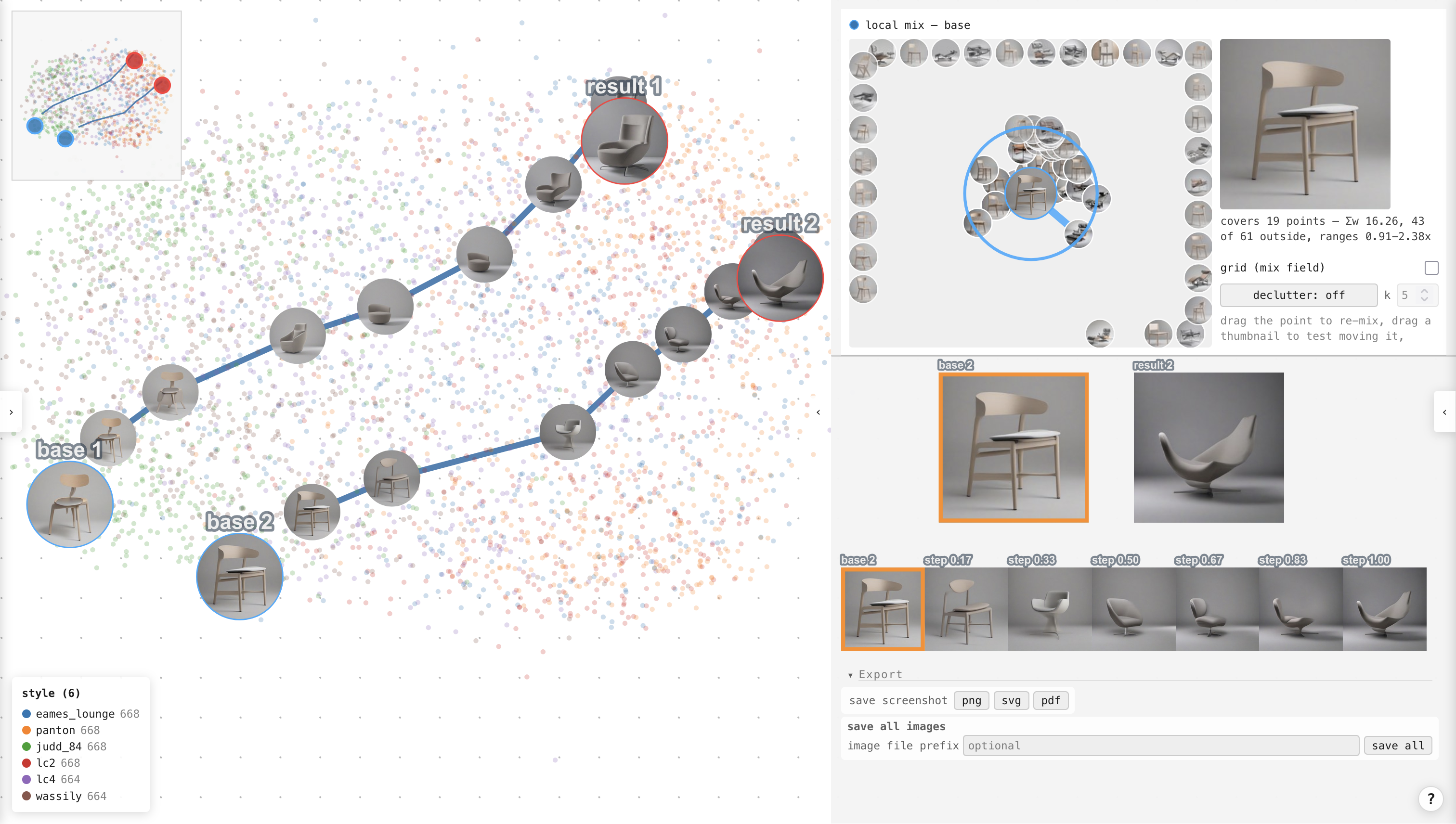}
    \caption{\sysname applied to a diffusion model for exploring chair styles (pairs mode).}
    \Description{A screenshot of \sysname in pair mode showing interpolation between diffusion-generated chair examples.}
    \label{fig:chair-2}
\end{figure}

\begin{figure}
    \centering
    \includegraphics[width=1\linewidth]{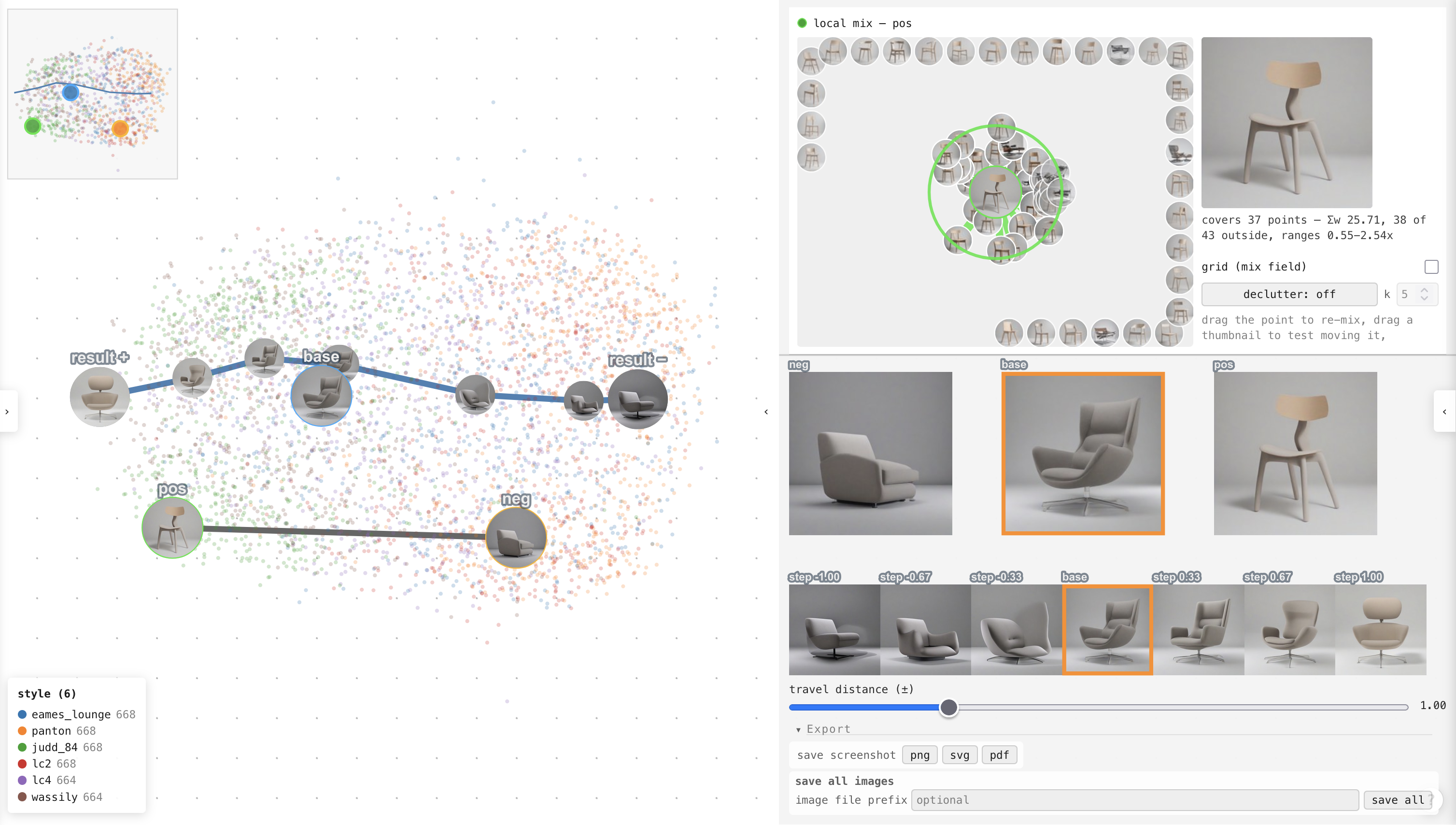}
    \caption{\sysname applied to a diffusion model for exploring chair styles (steering mode)}
    \Description{A screenshot of \sysname in steering mode showing chair-style variation defined by a base example and contrasting examples.}
    \label{fig:chair-3}
\end{figure}

The resulting map organizes chair latent codes primarily by style and material, with samples of the same style forming approximate clusters, as indicated by the color-coded points in \autoref{fig:chair-1}. Interpolation between points in the map and style manipulation using positive and negative examples are shown in \autoref{fig:chair-2} and \autoref{fig:chair-3}, respectively.

\subsubsection{Comments}
Due to hardware constraints, the \sysname interface operates only near real time for this use case. In the base-lens mode, each interaction requires denoising and decoding only a single sample, allowing relatively responsive operations such as dragging or resizing the lens. In pair and steer modes, however, an interaction may require decoding an entire interpolation or manipulation trajectory, resulting in several seconds of latency before denoising completes.

The primary purpose of this use case is to demonstrate that the interaction framework of \sysname generalizes beyond GANs to more recent generative architectures, including diffusion models with iterative and potentially stochastic generation processes. In future work, we plan to investigate alternative locations for interpolation and steering, including intermediate denoising states and internal model representations, rather than operating only on fully denoised latent codes.







\subsection{Case 3: Formative case study on Rotational Kirigami Systems}

\begin{figure}[t]
    \centering
    \includegraphics[width=1\linewidth]{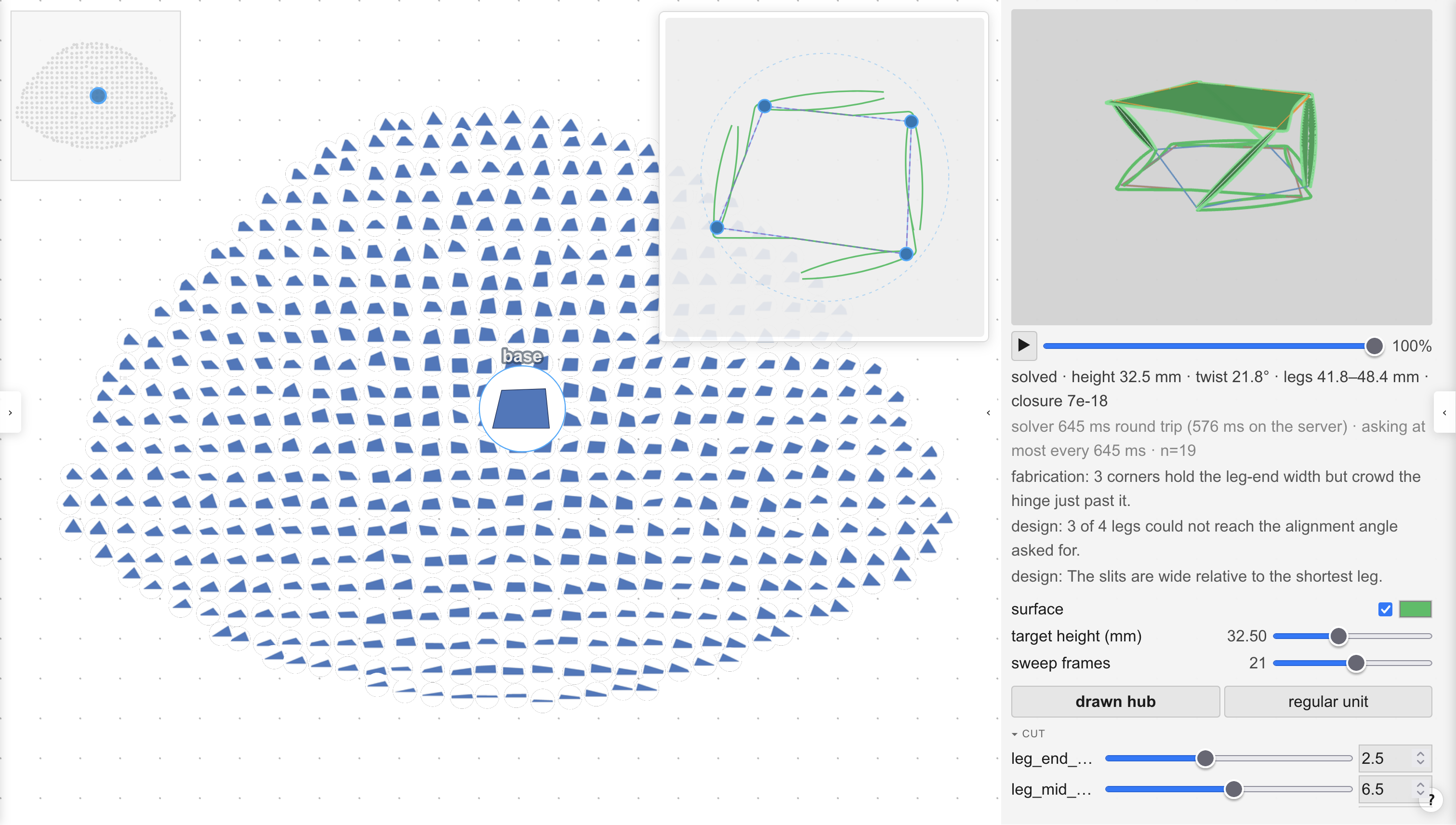}
    \caption{\sysname applied to kirigami unit design. The overview shows the sampled quadrilateral design space, while the local editor lets the user modify a selected hub polygon. Each edit is passed to the geometric solver, which updates the corresponding cut pattern, deployed unit, and feasibility checks.}
    \Description{A kirigami design interface with an overview of sampled quadrilateral designs and a local hub-polygon editor, alongside the corresponding cut pattern, deployed unit, and feasibility information.}
    \label{fig:kirigami-1}
\end{figure}

\paragraph{Rotational Kirigami Systems}
We use rotational kirigami as a formative engineering case study. The domain expert is an engineering researcher who has worked on rotational kirigami systems for five years. We therefore treat this case as a formative case study of \sysname on a problem from their research.

\begin{figure*}[t]
  \centering

  \begin{subfigure}[t]{0.236\textwidth}
    \centering
    \includegraphics[width=0.75\textwidth]{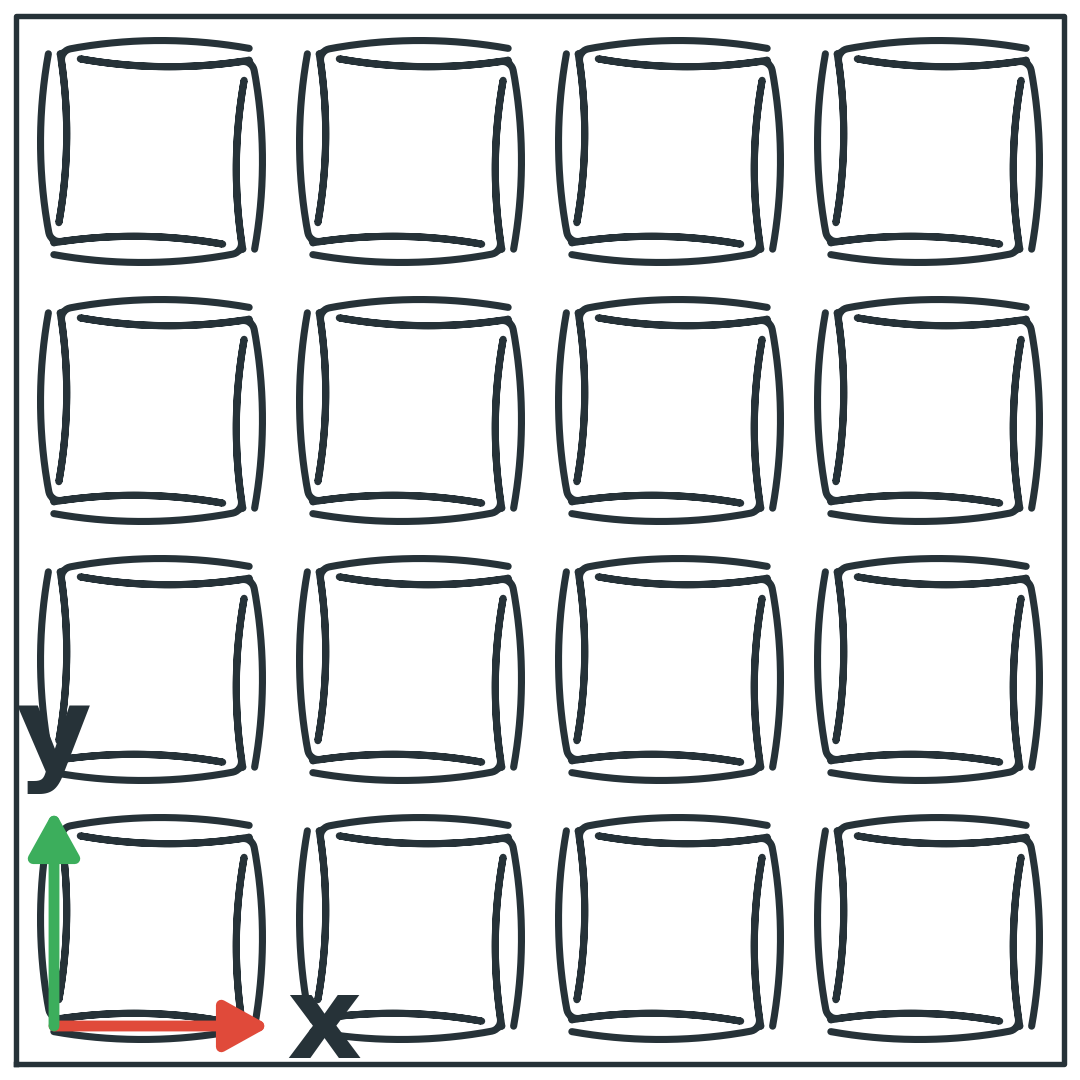}
    \caption{flat cut}
    \Description{A flat cut pattern for a kirigami sheet with hubs connected to the surrounding sheet by legs.}
    \label{fig:kirigami-a}
  \end{subfigure}\hfill
  \begin{subfigure}[t]{0.236\textwidth}
    \centering
    \includegraphics[width=\textwidth,keepaspectratio]{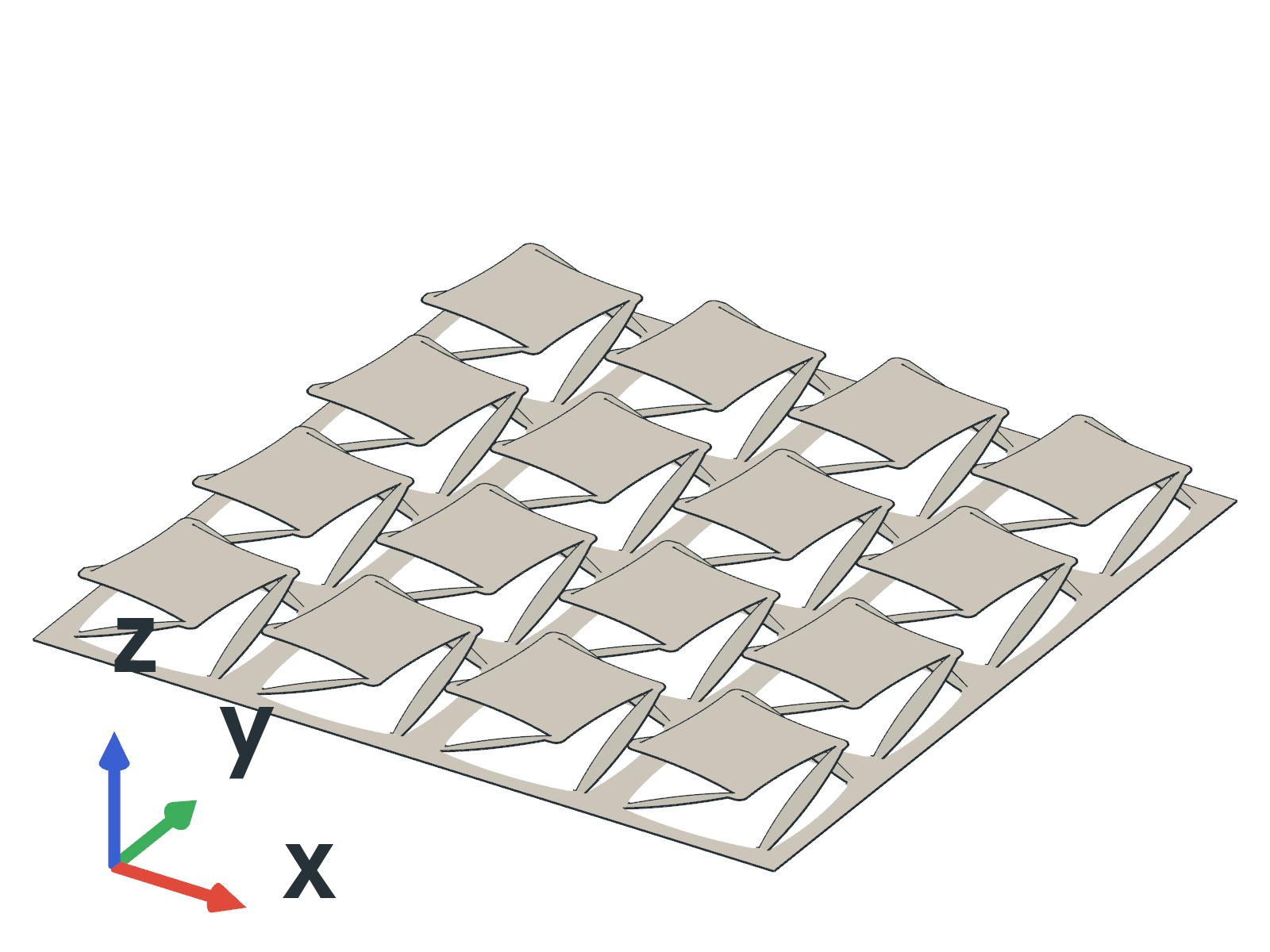}
    \caption{mid-deployment}
    \Description{A partially deployed kirigami sheet with hubs raised and rotated out of the sheet plane.}
    \label{fig:kirigami-b}
  \end{subfigure}\hfill
  \begin{subfigure}[t]{0.236\textwidth}
    \centering
    \includegraphics[width=\textwidth,keepaspectratio]{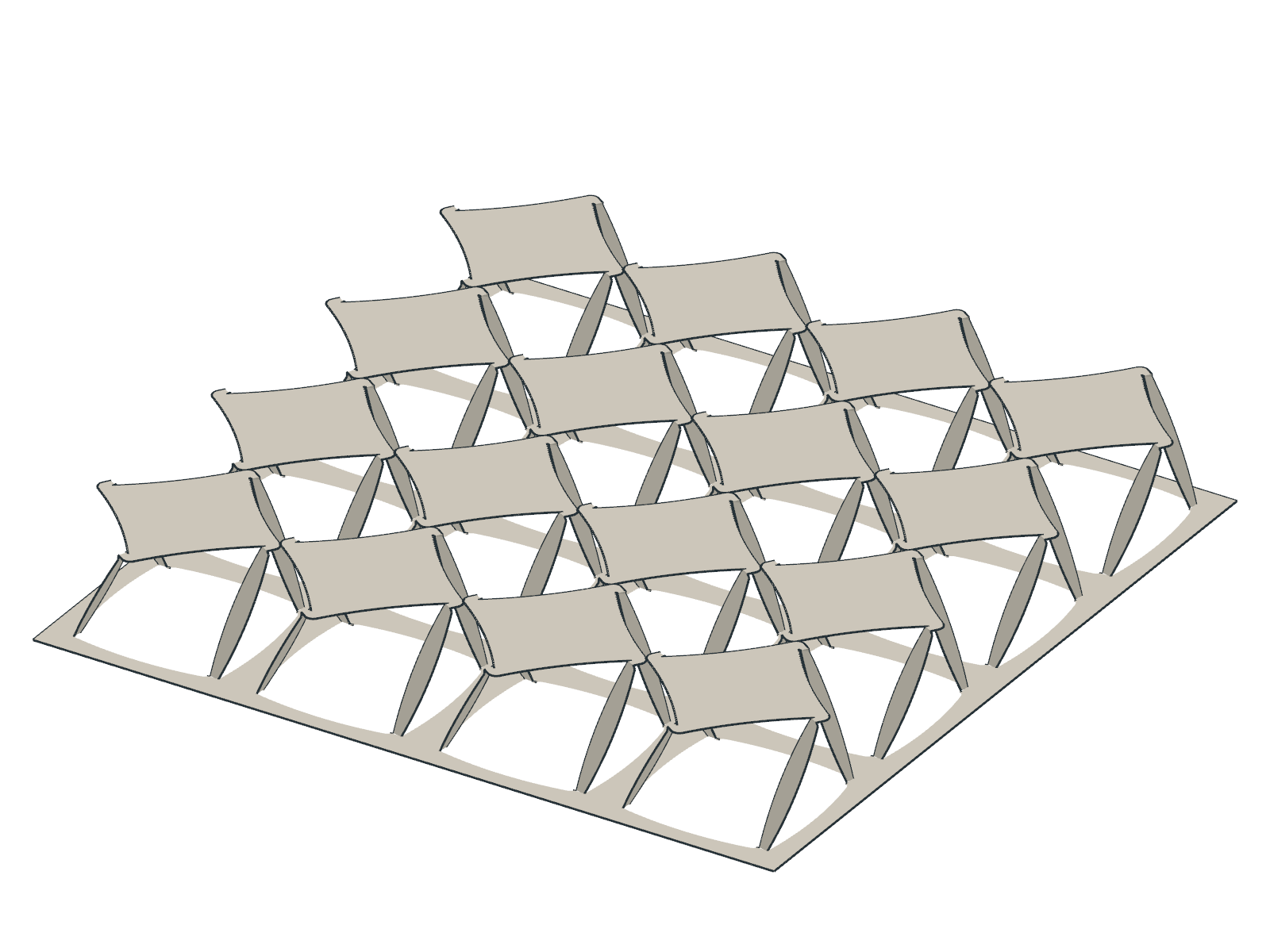}
    \caption{design height}
    \Description{A kirigami space frame at its design height, with neighboring hub corners meeting.}
    \label{fig:kirigami-c}
  \end{subfigure}\hfill
  \begin{subfigure}[t]{0.236\textwidth}
    \centering
    \includegraphics[width=\textwidth,keepaspectratio]{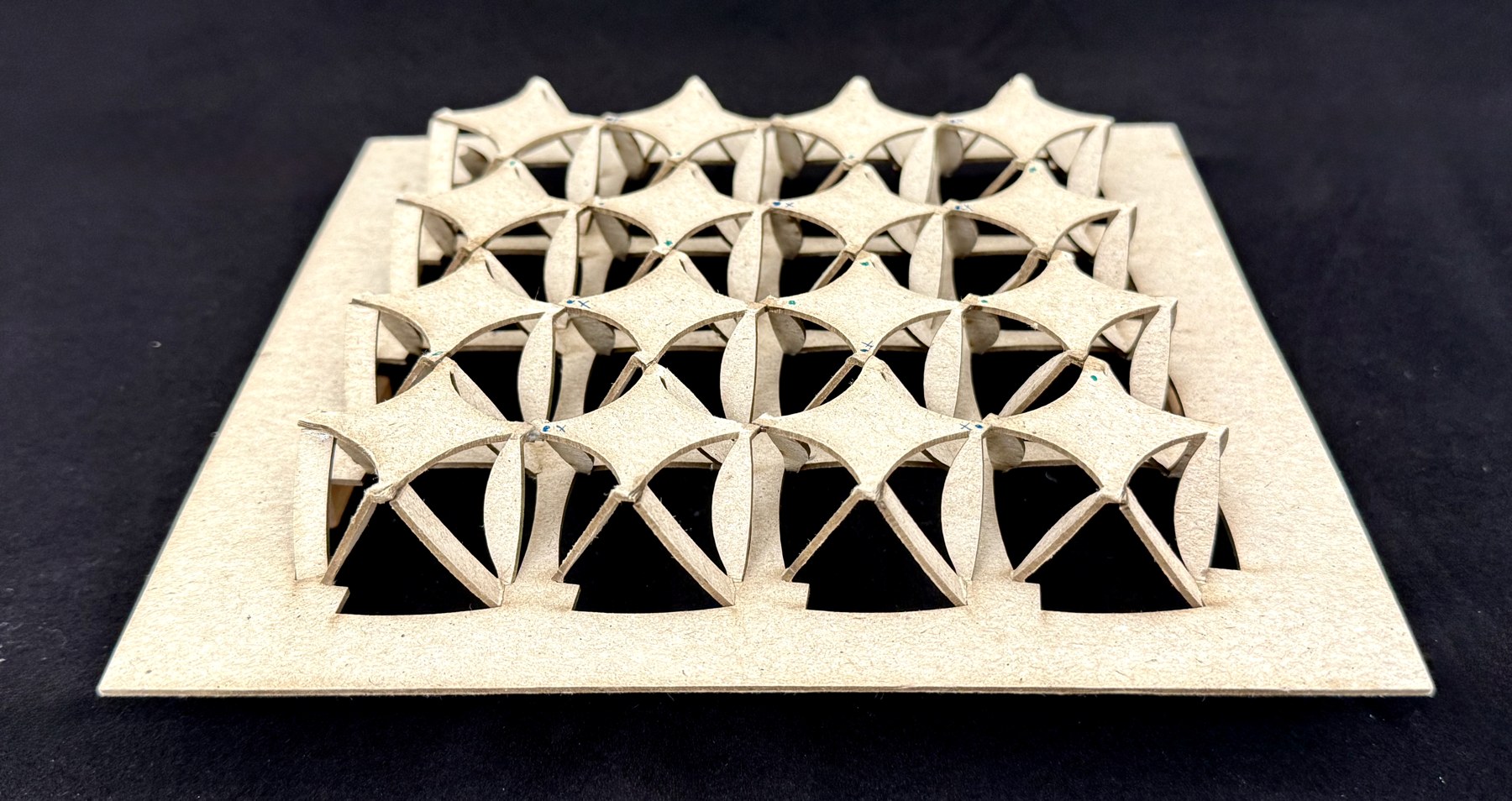}
    \caption{prototype}
    \Description{A photograph of a fabricated chipboard kirigami space-frame prototype in its deployed configuration.}
    \label{fig:kirigami-d}
  \end{subfigure}

  \begin{subfigure}[t]{0.236\textwidth}
    \centering
    \includegraphics[height=0.45\linewidth]{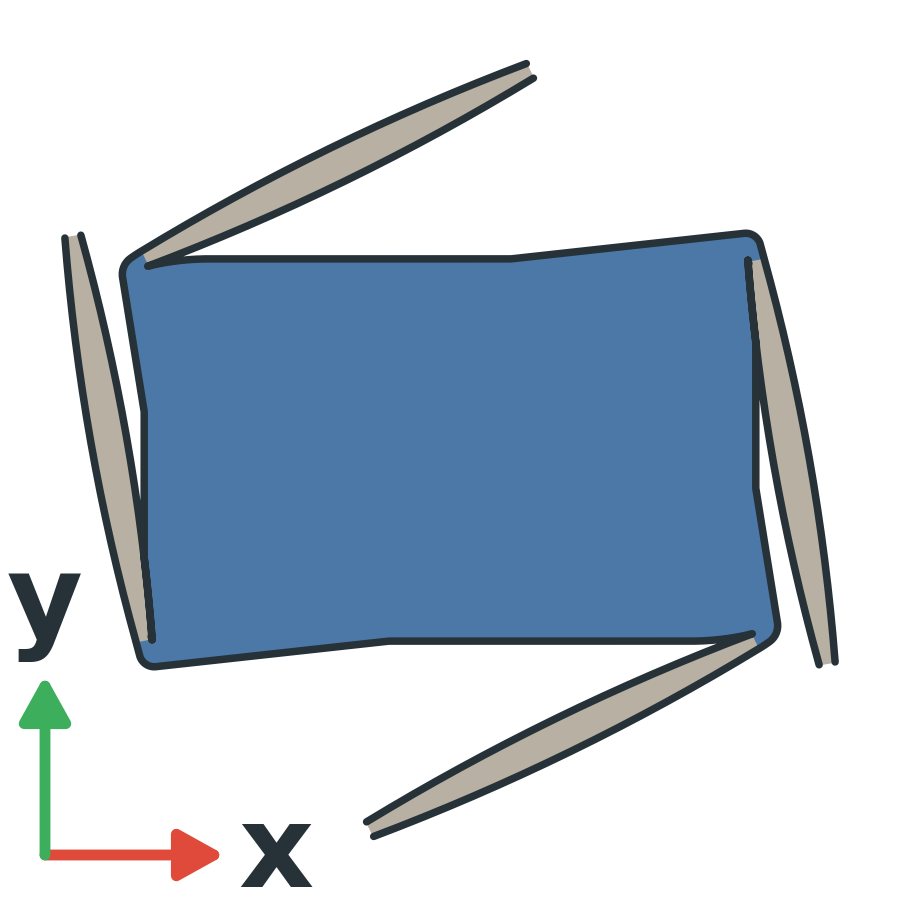}
    \includegraphics[height=0.45\linewidth]{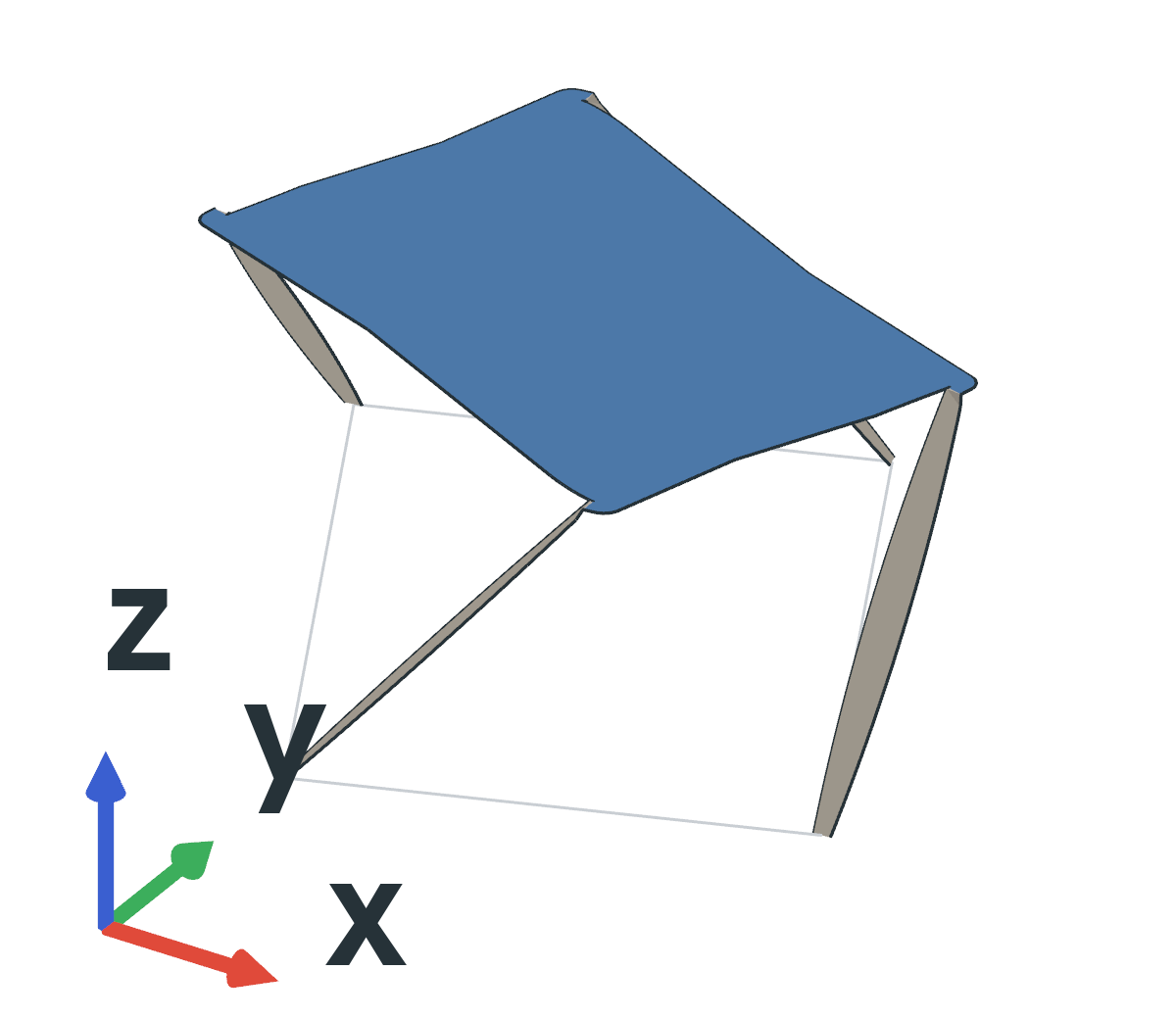}
    \caption{}
    \Description{A rectangular-hub kirigami unit, with its flat cut pattern on the left and deployed form on the right.}
    \label{fig:kirigami-e}
  \end{subfigure}\hfill
  \begin{subfigure}[t]{0.236\textwidth}
    \centering
    \includegraphics[height=0.45\linewidth]{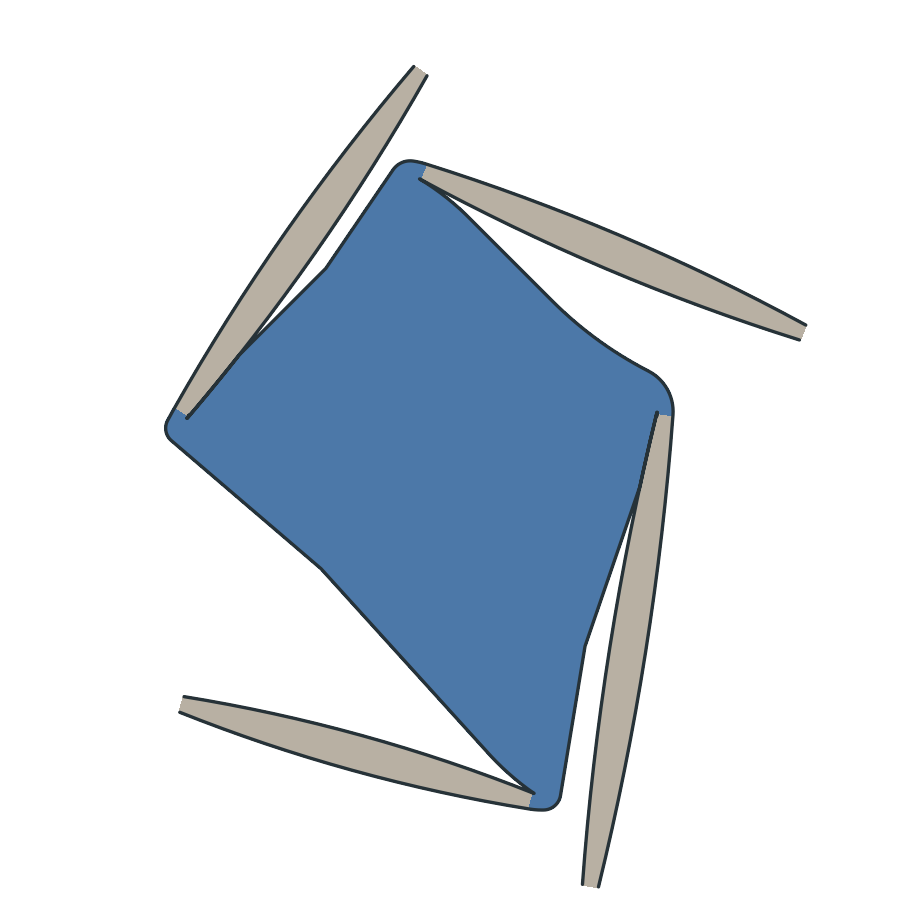}
    \includegraphics[height=0.45\linewidth]{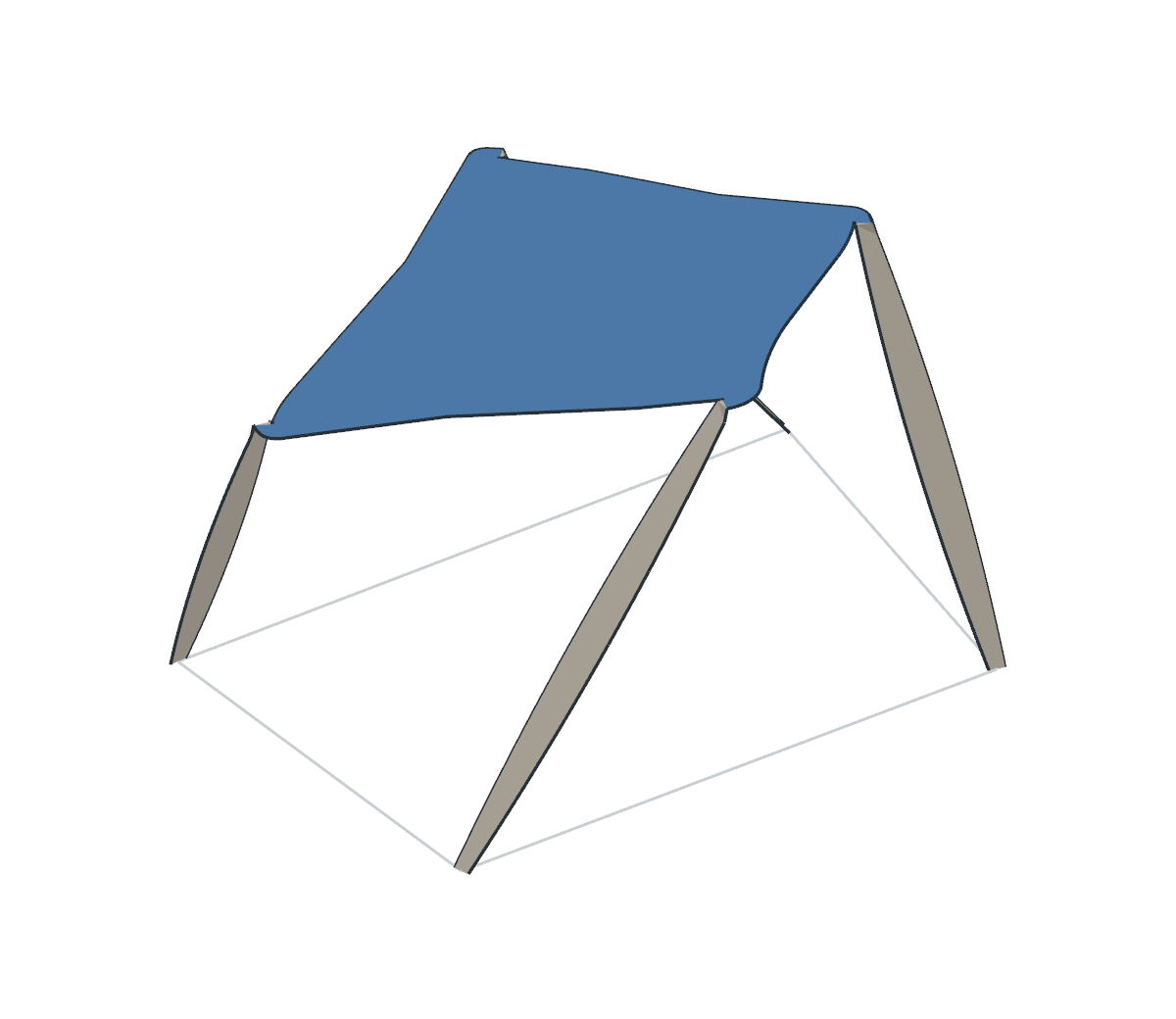}
    \caption{}
    \Description{A kite-shaped-hub kirigami unit, with its flat cut pattern on the left and deployed form on the right.}
    \label{fig:kirigami-f}
  \end{subfigure}\hfill
  \begin{subfigure}[t]{0.236\textwidth}
    \centering
    \includegraphics[height=0.45\linewidth]{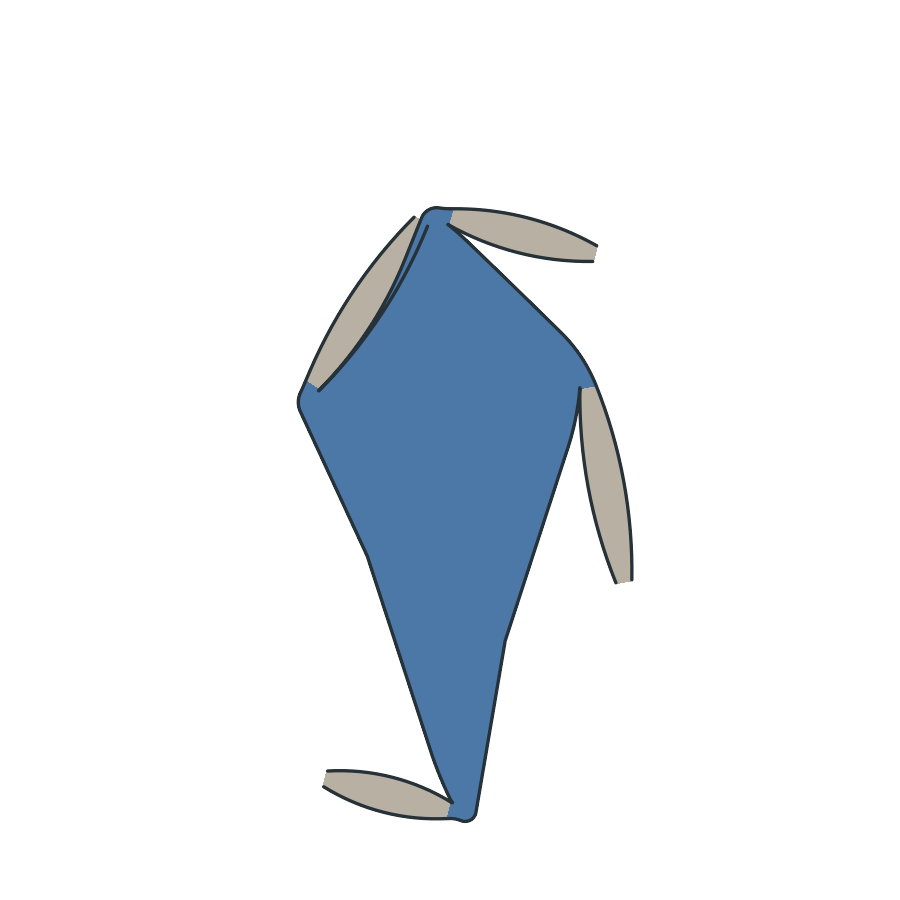}
    \includegraphics[height=0.45\linewidth]{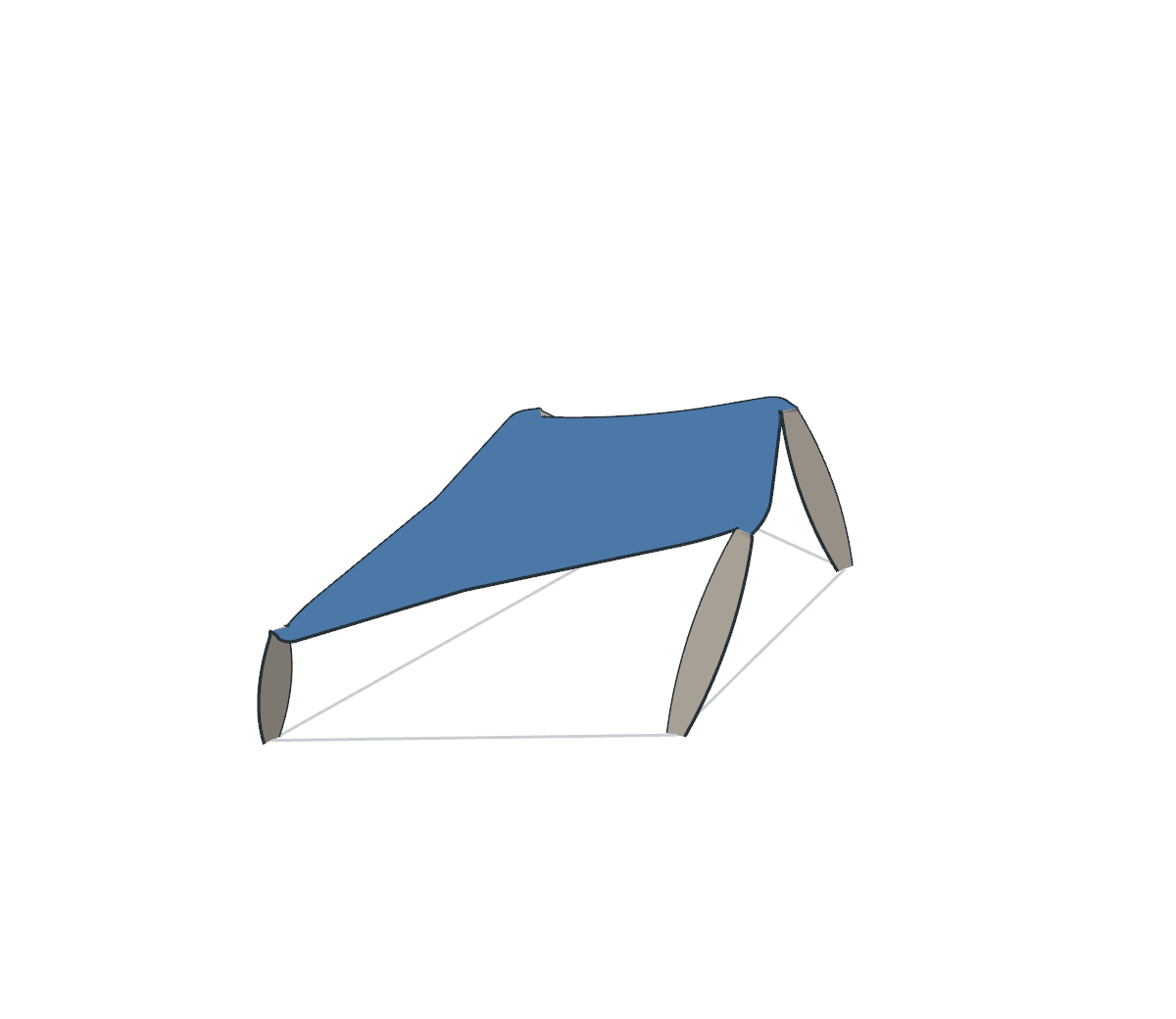}
    \caption{}
    \Description{A narrow kite-shaped-hub kirigami unit, with its flat cut pattern on the left and deployed form on the right.}
    \label{fig:kirigami-g}
  \end{subfigure}\hfill
  \begin{subfigure}[t]{0.236\textwidth}
    \centering
    \includegraphics[height=0.45\linewidth]{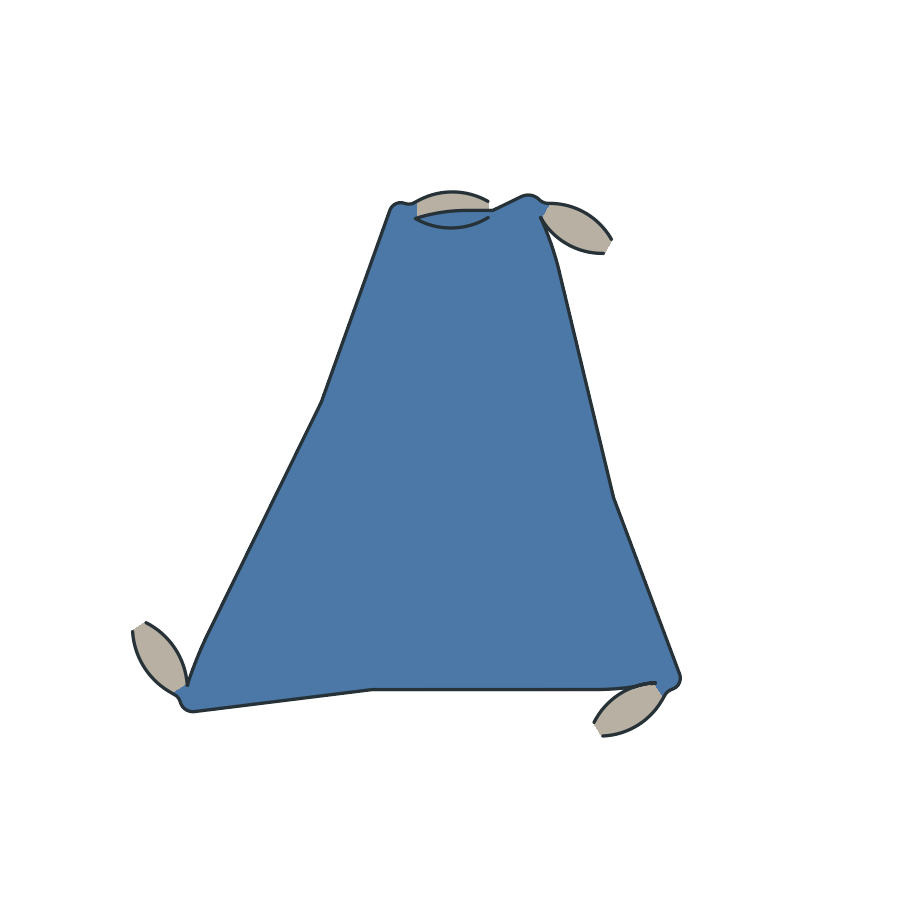}
    \includegraphics[height=0.45\linewidth]{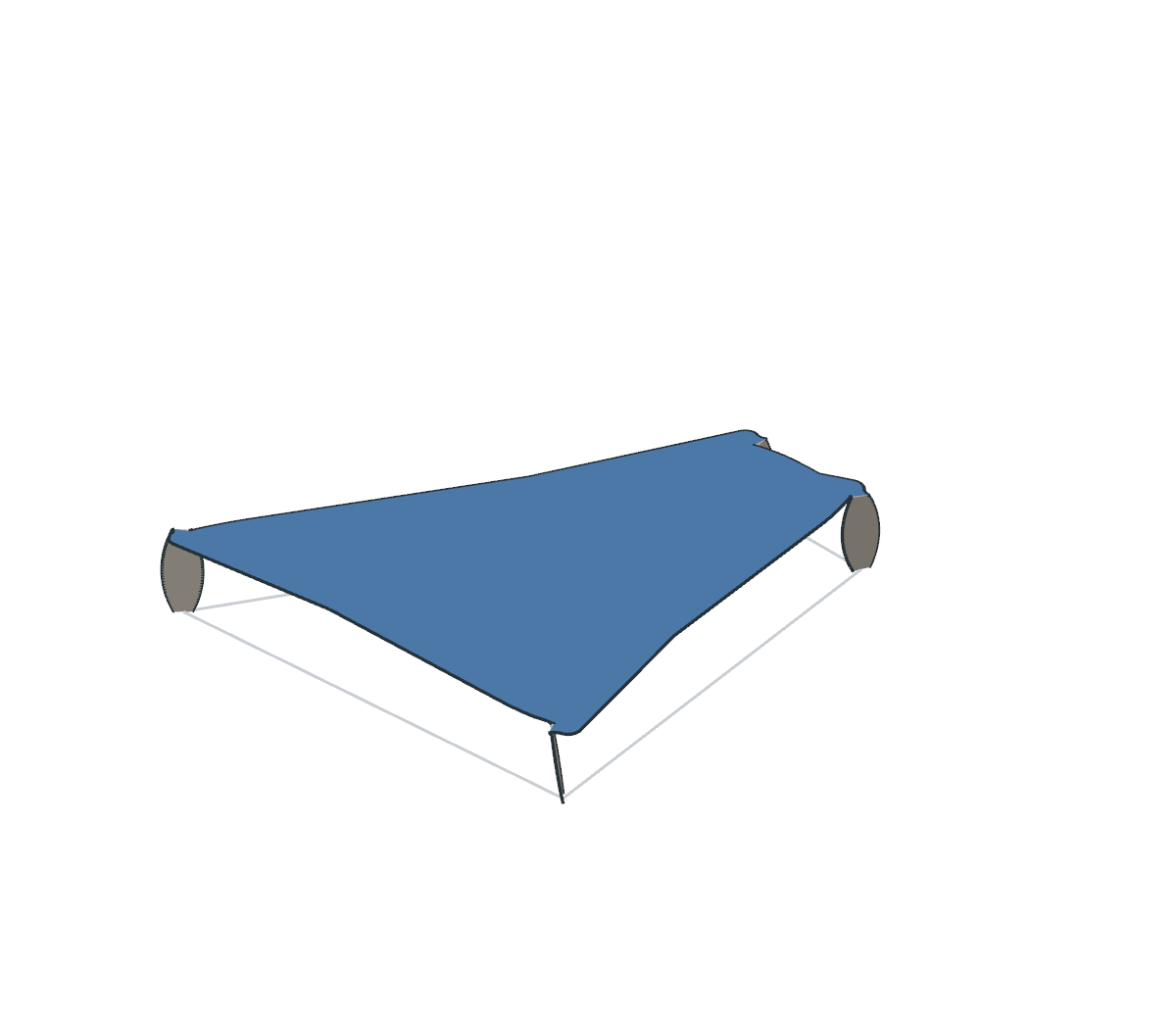}
    \caption{}
    \Description{A nearly triangular quadrilateral-hub kirigami unit, with its flat cut pattern on the left and deployed form on the right.}
    \label{fig:kirigami-h}
  \end{subfigure}

  \caption{Spin-valence kirigami from space frame to unit-level design. (a--d) A flat cut sheet deploys into a three-dimensional space frame: the hubs rise and rotate until neighboring hubs meet at the design height, where they can be joined to form the load-bearing structure. Panel (d) shows a fabricated chipboard prototype. (e--h) Different hub geometries produce different leg arrangements, cut patterns, and deployed units. These unit-level variations form the geometric design space explored in \sysname.}

  \Description{Top row: a flat kirigami cut pattern, an intermediate deployment, the space frame at its design height, and a fabricated chipboard prototype. Bottom row: four different hub geometries, each shown with its flat cut pattern on the left and corresponding deployed unit on the right.}

  \label{fig:kirigami-domain}
\end{figure*}

Spin-valence kirigami space frames are deployable structures cut from flat sheets~\cite{baker2014search,bakerSpinValenceKirigamiSpace2018,deoliveiraEffectGeometricTiling2024}. The cut pattern defines hub plates connected to the surrounding sheet by slender legs. During deployment, the hubs lift and rotate out of plane of the sheet until neighboring hubs meet and can be joined, transforming the sheet into a three-dimensional, two-way spanning space frame (\autoref{fig:kirigami-domain}a--d)~\cite{sherrow-grovesStructuralFiniteelementAnalysis2019,sahucExperimentalAnalyticalInvestigation2019}. Narrow regions at the ends of the legs localize deformation, enabling predictable deployment without prescribed fold lines. This allows structural materials such as steel sheets to be fabricated only by cutting, transported flat, and deployed with relatively low force. Once connected, tested structures sustained loads up to 20 times greater than the force required for deployment~\cite{deoliveira2023experimental}.

The cut geometry controls both the deployed form and the resulting structural behavior, allowing the same fabrication principle to produce structures with different shapes and mechanical properties~\cite{bakerSpinValenceKirigamiSpace2018,ramirezDoubleCurvedSpinValenceGeometric2020,deoliveira2023experimental,deoliveiraEffectGeometricTiling2024,XXVIIIGenerativeArt}. Related cut-and-rotate systems appear across scales in haptic interfaces, reconfigurable surfaces, optical metamaterials, and architectural skins~\cite{liuNanokirigamiGiantOptical2018,changKirigamiHapticSwatches2020,zhangProgrammableReconfigurableSurfaces2021,vazquezShapechangingArchitecturalSkins2019}. Their behavior is similarly governed by geometry, making the exploration of geometric design spaces relevant beyond the structural application considered here.

A space frame consists of repeated or varying kirigami units, and we apply \sysname at this unit level. For a prescribed deployment height, changing the hub polygon changes the leg geometry, cut pattern and deployed geometry (\autoref{fig:kirigami-domain}e--h). We therefore explore hub geometries and their feasibility before the subsequent problem of arranging different units into a compatible space frame, which is outside the scope of this case study. 

\subsubsection{Quadrilateral Dataset}
For this case study, we restrict the unit design space to quadrilateral hubs. We randomly generated 20,000 convex quadrilaterals using Valtr's algorithm~\cite{Valtr1995}. Each polygon is represented by the 2D coordinates of its four vertices, yielding an 8-dimensional vector representation, although the intrinsic degrees of freedom are lower due to geometric constraints and symmetries. We canonicalize each polygon to account for translation and cyclic re-indexing of its vertices by ordering the vertices counterclockwise and rotating the polygon such that its first edge is aligned with the positive x-axis. As in the other use cases, we use UMAP to project this 8D representation of the configuration space into two or three dimensions for visualization and interactive exploration in \sysname.


\subsubsection{Geometric Model and Leg Placement Solver}
Unlike the previous use cases, the kirigami example does not use a learned generative model. Each point in the design space directly represents a hub polygon through its vertex coordinates.
We connect this representation to a leg-placement solver that converts a selected hub polygon into a deployable kirigami unit (\autoref{fig:kirigami-domain}e--h).

The polygon defines the hub plate, with one leg attached at each vertex. Given a target deployment height and hub rotation, the solver determines compatible leg endpoints while preserving leg length through deployment. It then constructs the cut pattern and checks geometric and fabrication constraints, including intersecting cuts, severed material regions, and insufficient hinge widths. After each vertex edit, the solver recomputes the cut pattern, deployed unit, and feasibility checks.

This connection between interactive geometry and the solver makes local refinement specific to the engineering task. 
Rather than interpolating between nearby examples, a designer directly edits a selected polygon while the system recomputes the corresponding kirigami construction. The overview therefore supports broad navigation across hub shapes, while the polygon editor supports refinement of a selected design.


\subsubsection{Interview Process} 
We conducted a recorded 40-minute formative design walkthrough in \sysname with the domain expert. After introducing the main navigation interactions, the expert freely explored the 2D and 3D projections, navigated the quadrilateral design space, and refined selected geometries using the polygon editor and deployment controls. The exploration also included attempts to recover feasible designs when the solver could not satisfy fabrication constraints. The session concluded with a discussion of useful interactions, points of difficulty, and additional information or controls needed for structural design.

\subsubsection{Findings}


\paragraph{Interface: Global Navigation and Local Refinement}
The global exploration map changed how the expert interacted with the design space. Rather than modifying vertices or deployment parameters one at a time, the expert could move through the map and rapidly test substantially different polygon geometries. This made broad exploration more natural and encouraged consideration of configurations that would otherwise require several manual edits. Once an interesting configuration was identified in the global map, the interaction shifted toward local refinement through direct vertex editing, thus supporting a two-stage workflow. 

The 2D and 3D projections supported different aspects of this exploration. The expert initially found the 2D projection easier for forming a mental map of the design space and recognizing geometric regions. For example, some boundary regions contained nearly collapsed quadrilaterals approaching triangular shapes, whereas other regions contained more evenly proportioned quadrilaterals, including square-like shapes and trapezoids (\autoref{fig:kirigami-1}). The 3D view was initially harder to interpret, but became more useful once the 2D organization was understood, providing additional spatial context and finer control over changes such as varying the length of one side. This suggests that global navigation may benefit from multiple complementary representations: a simpler view for orientation and a higher-dimensional view for more detailed exploration.

For local refinement, the expert found direct vertex editing more appropriate than interpolation among neighboring examples. Once an interesting region had been identified, the relevant question shifted from finding nearby polygons to modifying the candidate while preserving a feasible kirigami construction. Because the polygon editor was coupled to the geometric solver, moving a vertex changed the cut pattern and deployed geometry and could also make the unit infeasible. Local refinement therefore became an interaction with the constraints of the physical design.

\paragraph{Interface: Constraint Awareness and Relaxation Guidance}
The walkthrough also highlighted the importance to support constrained exploration. When the solver could not construct a feasible unit, the interface reported infeasibility but provided little guidance on how to recover. The expert responded by adjusting the hub's target height and twist angles and by modifying polygon vertices until feasibility was restored, with limited information about which change would be most effective. Small vertex edits could often recover a feasible solution, 
but the recovery process still relied largely on trial and error.

For constrained design spaces, feasibility can therefore serve not only as a property to report, but as active guidance for navigation and refinement. The expert suggested several complementary forms of guidance: (1) globally indicating where nearby feasible designs lie, (2) suggesting local directions in shape geometry that move the current design toward feasibility, and (3) recommending how constraints such as target height or allowable twist might be relaxed. These interactions distinguish between navigating the current feasible region and modifying the constraints that define it.

\paragraph{Domain-Specific Extensions for Kirigami Design}

The walkthrough also showed that geometric similarity is only one way to organize a kirigami design space. A geometry-based map was useful for discovering polygon families, but engineering decisions also depend on the consequences of that geometry, including deployed height and shape, stiffness, buckling behavior, and hub rotation under load, which could be used to color, filter, or reorganize the overview. A deployed-geometry view could likewise show the physical consequences of a design directly in the map, while the vertex editor could expose alternative leg placements that remain feasible under the current constraints. These additions would move the interface from browsing geometric variation toward selecting and refining designs based on their intended behavior, first at the unit level and ultimately for the complete kirigami space frame.



\section{Discussion and Future Work}\label{sect:discussion}

\subsection{Characteristics of interpolation}

Fixing the anchor points yields only a fixed 2D parameterization and therefore cannot, in general, provide full navigation of the latent space. To characterize the maximal coverage of the interpolation scheme, we instead consider a hypothetical setting in which users are free to place and reposition the anchors, potentially through careful and time-consuming manual adjustment. Because the interpolation weights are normalized and nonnegative, any interpolated point lies in the convex hull of the $K$ anchor points, which is a $(K-1)$-simplex when the anchors are affinely independent. Without normalization, nonnegative weighted combinations would instead span the convex cone generated by the anchors. For any fixed anchor configuration and weighting function, however, sweeping the 2D interface traces only a 2D subset of this convex hull. We therefore consider the \emph{coverage} of this induced 2D surface at several levels: its relationship to the anchor-induced convex hull, to the ambient latent space, and ultimately to the semantic variation of the generated outputs reachable through the interpolation.

Here, we highlight several desirable properties achieved by our interpolation design. Given a set of latent space anchor points and their corresponding 2D placements, sweeping the control point across the canvas and applying the weighted interpolation at each location induces a 2D parameterization of the high-dimensional latent space. The properties of this parameterization depend jointly on 1) the weighting function, 2) the choice of latent space anchors, and 3) their 2D layout. We characterize this embedding map and its induced surface into some desirable properties of an interpolation scheme. 
In the next subsection of the Discussion (\autoref{sect:discussion}), we further consider inverse-design approaches that directly optimize anchor layouts and weighting functions with respect to these and additional desirable properties.

\paragraph{Coverage.}
Movable reference points allow the interface to realize convex combinations within their simplex, preserving reachability of the corresponding convex region in latent space.

\paragraph{Continuity and Smoothness}
Continuous weighting functions induce continuous navigation trajectories, while differentiable weights produce smoother navigation surfaces and reduce abrupt changes in the generated output. Among the three weighting functions considered in \autoref{eq:weight-function-1} - \ref{eq:weight-function-3}, all yield continuous interpolation, while the negative-log function (\autoref{eq:weight-function-3}) with a cutoff is not differentiable at the cutoff radius. As a result, when a neighboring anchor crosses this boundary, its influence drops to zero with an abrupt change in rate, which can appear as a sudden change in the ``velocity'' of the generated output. From the perspective of the 2D-to-latent mapping, sweeping over the interface for a fixed reference-point layout induces a 2D surface in latent space; non-differentiable weighting functions can introduce creases into this surface.

\paragraph{Reference-Point Interpolation.}
Inverse-distance weighting approaches exact interpolation as the control point approaches a reference point. Smoothstep weighting may not, depending on its support and normalization.

\paragraph{Dimension Sparsity.}
Finite-support weights restrict mixtures to nearby reference points, reducing local effective dimensionality. This can support sparse representations and layouts that place correlated or co-activated latent features near one another.

\subsection{Inverse Design of Local Refinement Layout}

In this work, we manually designed the interpolation and parameterization scheme underlying the interaction, extending the intuitive notion of color mixing to high-dimensional space navigation. 
We demonstrate that this design satisfies several desirable properties, including smoothness and continuity, as well as geometric constraints such as interpolation through specified anchor points.
However, manually designing an interpolation scheme can introduce implicit trade-offs or incompatibilities among these properties, and may overlook properties that are important for particular models, tasks, or interaction contexts.
As the intrinsic dimensionality, or degrees of freedom, of the design space increases, a single 2D or 3D local layout can no longer capture all meaningful variation. How to factorize these variations and map them into a coherent set of navigable views therefore becomes an important interface-design challenge.

In future work, we would consider formulating this problem in the inverse direction: rather than manually specifying the mapping, we could derive the low-to-high-dimensional interface-to-latent parameterization from an explicit set of design desiderata. For example, what neighborhood layout is most appropriate for controlling generative models through sparse autoencoder representations? How should the interface-to-latent mapping be reparameterized so that points generated through the interface remain consistent with their locations when projected back into the global overview? Is there a combination of neighborhood layout, weighting function, and interpolation scheme that produces a more perceptually uniform landscape in the interface? Such formulations would make the trade-offs among mapping properties explicit, allow different properties to be weighted according to task requirements, and enable more automated construction of interaction mappings tailored to different models and use cases.

\subsection{Decoupling Latent Space Structure from Navigation}
So far, the simplest form of the exploration interface combines a 2D UMAP projection with a 2D mixing palette. However, our formative study suggests that users benefit from both 2D and 3D representations of the UMAP embedding. This distinction points to two related but separable tasks: understanding the structure and geometry of the latent space distribution, and navigating the generative space through interaction. By decoupling these functions, the interface can be reorganized so that dimensionality-reduction views support structural exploration, while the mixing palette provides a consistent space for navigation and steering.

\subsection{Boundary Conditions and Constraints}

Our examples assume unrestricted navigation, but not every location in representation space yields a meaningful, valid, or desirable output. Safety requirements, physical laws, fabrication limits, and domain-specific constraints may restrict the regions users should explore.

Navigation could remain near a valid-data manifold or use model density as a soft boundary to discourage movement far from the training distribution. Interfaces could incorporate these constraints by projecting movements onto valid regions, adjusting steering directions near boundaries, or indicating when trajectories approach infeasible areas. Although beyond this article's scope, constraint-aware navigation is an important step toward practical steering interfaces.

\subsection{Non-Euclidean Curved Spaces}

Our vector arithmetic assumes a locally Euclidean latent space, where steering always follows straight-line trajectories. In practice, valid model outputs may lie on a curved manifold (\autoref{sect:related-work}). Straight-line interpolation can then leave this manifold, pass through low-density ``voids,'' and produce poor-quality, meaningless, or unpredictable outputs. Likewise, a direction that is meaningful locally may not remain semantically consistent elsewhere. In such spaces, navigation may therefore require locally varying directions or \textit{manifold-following paths} rather than a single global steering vector.

\subsection{Beyond Latent Space}
The same geometric intuition extends beyond latent representations. Some steering methods modify the model itself rather than moving individual samples through a fixed latent space. Low-Rank Adaptation (LoRA), for example, represents a compact update to model parameters that can be scaled, combined, or contrasted with other updates. Systems such as SliderSpace expose these learned modifications through continuous controls, making parameter-space steering conceptually similar to adjusting the step size $\alpha$ along a latent direction.

This suggests that the framework of \texttt{base} + \texttt{direction} + \texttt{step size} applies more broadly than latent space editing. The base may be a latent code, activation state, parametric model input, or model parameters; the direction may be a learned semantic vector, activation difference, or low-rank parameter update; and the step size controls the strength of the intervention. From an interface perspective, these different mechanisms can be expressed through a common visual language.

\section{Conclusion}

We presented \sysname, an interaction framework and visual interface for steering generative models through coordinated global exploration and local refinement. Rather than requiring users to begin with a well-specified editing goal, \sysname allows them to discover useful contrasts from examples, construct steering operations from a base point, direction, and step size, inspect the resulting trajectories, and refine the examples that define those operations. Its linked global overview and local mixing view support this process at complementary scales while preserving a consistent spatial mental model.

Across StyleGAN2, Stable Diffusion XL, and a rotational-kirigami design space, we showed how the framework can support both generative-model exploration and domain-specific uses. These examples also expose important limitations and opportunities: steering directions may only be locally valid, diffusion models introduce substantial interaction latency, and constrained design spaces require stronger feasibility guidance. More broadly, our results suggest that generative-model interfaces can benefit from treating steering not only as parameter adjustment, but as an iterative process of navigating, inspecting, and revising transformations across a structured generative landscape.

\bibliographystyle{ACM-Reference-Format}
\bibliography{bib,genaisurvey,genaigeometry}










\end{document}